\documentclass{article}

\usepackage{amsmath, amssymb}
\usepackage{graphics}
\usepackage{epsfig}

\newcommand{\e}{\mathrm{e}}
\newcommand{\eps}{\varepsilon}
\newcommand{\ie}{\textit{i.e.}}
\DeclareMathOperator{\tr}{tr}

\begin{document}

{\centerline{\LARGE{\textbf{Correlated random walks with a finite memory range}}}}

\vspace{0.5cm}

{\centerline{\large{Roger Bidaux\dag\ and Nino Boccara\dag\ddag}}}

\vspace{0.5cm}

{\centerline{\dag DRECAM/SPEC CE Saclay, 91191 Gif-sur-Yvette Cedex, France}}

\vspace{0.2cm}

{\centerline{\ddag Department of Physics, 
University of Illinois at Chicago, Chicago}}

\vspace{0.3cm}

{\centerline{email: {\texttt{bidaux@spec.saclay.cea.fr 
{\textrm{and}} boccara@uic.edu}}}}

\vspace{1cm}

{\textbf{Abstract.}
} We study a family of correlated one-dimensional 
random walks with a 
finite memory range $M$. These walks are extensions of the Taylor's walk
as investigated by Goldstein, which has a memory range equal to one. 
At each step, with a probability $p$, the random walker moves either 
to the right or to the left with equal probabilities, or with a probability 
$q=1-p$ performs a move, which is a stochastic Boolean function of the 
$M$ previous steps. 
We first derive the most general form of this stochastic Boolean function, 
and study some typical cases which ensure that the average value $<R_n>$ 
of the
walker's location after $n$ steps is zero for all values of $n$. 
In each case, using a matrix technique, we provide a general method for 
constructing the generating function of the probability distribution of 
$R_n$; we also establish directly an exact analytic expression for the
step-step correlations and the variance $<R_n^2>$ of the walk. From the 
expression of $<R_n^2>$, which is not straightforward to derive from 
the probability distribution, we show that, for $n$ going to infinity,
the variance of any of these walks behaves as $n$, provided $p>0$. 
Moreover, in many cases, for a very small fixed value of $p$, the variance 
exhibits a crossover phenomenon as $n$ increases from a not too large value.
The crossover takes place for values of $n$ around $1/p$. This feature may 
mimic the existence of a non-trivial Hurst exponent, and induce a misleading 
analysis of numerical data issued from mathematical or natural sciences
experiments.

\newpage

\section{Introduction.}

Soon after its early developments in probability theory [1] the study 
of random walks and their applications has gradually carved its way into 
the concerns of many topics in exact or applied science. In various 
recent developments, the notion of ``walk'' in the proper sense of the 
term has gradually blurred, giving way to a more general concept where 
a ``step'' of the walk is not necessarily an algebraic distance between 
two locations but may become a time interval between two consecutive
events of a chronological series, or even a sequence of real numbers.

Before mentioning a few significant fields of application, let us quote 
more specifically the notion of ``correlated random walk,'' which underlies 
the possibility for a step to depend on the values taken by some (or all) of
the previous steps. We will address the step correlations and their 
consequences at length in the course of this paper, and will consider 
therefore the ``direct'' problem, in contradistinction with the usual concern
prevailing in applied science which asks for answers to the ``inverse'' problem,
\ie, can a given set of data be suitably described by a (correlated or 
uncorrelated) random walk formalism? Examples of fields resorting to the 
inverse problem are neuromuscular [2] and cardiovascular medicine [3], 
structural genetics [4-6], physiology [7], regulation of biological 
rhythms [8]; to these fields pertaining essentially to the medical domain can be 
added the soaring irruption of financial analysis [9-11] in view of market 
predictions.

In this paper we will focus on the formalism of correlated random walks on  
one-dimensional lattices and discuss the r\^ole of memory in some typical models 
which can be solved exactly. With the help of a direct evaluation of the variance 
of the walks, we will be able to prove that all walks with short-range memory 
have a Hurst exponent equal to $\frac{1}{2}$  in the strict limit of an infinite
number of steps, although the apparent Hurst exponent may look different from 
$\frac{1}{2}$. In particular, we will endeavor to produce a simple argument for 
a necessary condition yielding a Hurst exponent larger than $\frac{1}{2}$. 
 
\section{Correlated walks and memory}

The transition probabilities in a simple random walk on a 
one-dimensional infinite lattice are constant and independent of the
course of the walk. If these transition probabilities depend upon the 
$M$ previous steps, we have a correlated random walk whose memory range 
is $M$. Correlated random walks were introduced by Taylor [12] in an analysis of
diffusion by continuous motions. The Taylor's model is a one-dimensional 
random walk on a lattice, in which  steps can be made to nearest neighbors 
only. The walker has a probability $\alpha$ to repeat his previous move, 
and a probability $1-\alpha$ to move in the opposite direction. The resulting 
correlated random walk has a memory range equal to 1. If 
$\alpha > \frac{1}{2}$ (resp. $\alpha < \frac{1}{2}$) the 
walk is said to be {\it persistent} (resp.  {\it antipersistent}). A complete 
analysis of the Taylor's model has been given by Goldstein [13]. More examples 
of persistent random walks have been studied by many other authors [14-16]. 
A detailed bibliography may be found in Weiss [17]. 

In this paper we study a class of correlated random walks with memory 
range $M>1$. Our model may be described as follows. Let $(S_k)$ be 
an infinite sequence of random variables such that, for $k>M$,
$$
S_k = 
\begin{cases} 
1 & \text{with probability $\frac{1}{2}p$}\\
-1 & \text{with probability $\frac{1}{2}p$}\\
f_M(S_{k-1}, S_{k-2},\ldots,S_{k-M}) & \text{with probability $q=1-p$,}\\
\end{cases}
$$
where $f_M$, called the {\it memory function}, is a Boolean function from 
$\{-1,\!1\}^M$ onto $\{-1,\!1\}$.
For $k \leq M$, the $S_k$'s are independent, identically distributed (\textit{iid}) 
Bernoulli random variables, each 
$S_k$ taking the values $\pm 1$ with equal probabilities. The values of these 
$M$ random variables define the initial conditions for the walk. The position 
of the random walker after $n$ steps is
$$
R_n = S_1 + S_2 + \cdots + S_n.
$$
The set $\mathcal{F}_M$ of all bounded functions defined on $\{-1,1\}^M$ is a 
Euclidean space of dimension $2^M$ for the dot product defined by
\begin{eqnarray*}
\lefteqn{<f_M^{(1)}\mid f_M^{(2)}>\;\; = }\\ 
& & \left(\frac{1}{2}\right)^M
\tr f_M^{(1)}(S_{k-1}, S_{k-2},\ldots,S_{k-M})
f_M^{(2)}(S_{k-1}, S_{k-2},\ldots,S_{k-M}),\\
\end{eqnarray*}
where the trace operator $\tr$ is a sum over all the possible values of 
the $M$ variables $S_{k-1}, S_{k-2},\ldots,S_{k-M}$. 
The set of the following $2^M$ functions 
$$\left\lbrace
\begin{array}{l}
1,\\
S_{k-1}, S_{k-2}, \ldots , S_{k-M},\\
S_{k-1}S_{k-2}, S_{k-1}S_{k-3}, \ldots , S_{k-M+1}S_{k-M},\\
S_{k-1}S_{k-2}S_{k-3}, \ldots , S_{k-M+2}S_{k-M+1}S_{k-M}, \\ 
\cdots,\\
S_{k-1}S_{k-2}\cdots S_{k-M},\\
\end{array}
\right.$$
is a complete orthonormal system, and any function in $\mathcal{F}_M$ can be 
written as a linear combination of these $2^M$ functions. The most general
correlated random walk with a finite memory of range $M$ is, therefore, 
defined by a sequence $(S_k)$ of random variables such that, 
for $k>M$,
\begin{equation}
S_k = 
\begin{cases}
1 & \text{with probability $\frac{1}{2}p$}\\
-1 & \text{with probability $\frac{1}{2}p$}\\
\eps_{j_1j_2\cdots j_r}S_{k-j_1}S_{k-j_2}\cdots,S_{k-j_r}
& \text{with probability $qa_{j_1j_2\cdots j_r}$,}
\end{cases}
\label{general-model}
\end{equation}
where $p+q=1$, $1\leq j_1<j_2<\cdots<j_r\leq M$, and  
$\eps_{j_1j_2\cdots j_r}$, which are \textit{non-random}, are equal either to 
$1$ or to $-1$. The $a$'s are conditional probabilities (some of them possibly
equal to zero) satisfying the completeness relation: 
\begin{eqnarray*}
&  & \sum_{1\leq j_1\leq M}a_{j_1} + \sum_{1\leq j_1<j_2\leq M}a_{j_1j_2}
+\cdots\\
& & \qquad + \sum_{1\leq j_1<j_2\leq<j_r\leq M} a_{j_1j_2\cdots j_r}
+ \cdots + a_{12\cdots M} = 1.\\
\end{eqnarray*}

As above, for $k \leq M$, $S_k$ is a simple Bernoulli random variable 
which takes the values $\pm 1$ with equal probabilities.

In order to simplify this model, we require that, as in the case of 
a simple symmetric random walk, the average value $<R_n>$ of the position 
of the random walker after $n$ steps should be equal to zero. While this 
condition is automatically satisfied if $n\leq M$, a tedious inspection 
shows that, for $n>M$, the above requirement is satisfied if, and only if, 
the memory function $f_M$ has the following simple forms:

\begin{itemize}
\item $f_M(S_{k-1}, S_{k-2},\ldots,S_{k-M})$ is a linear combination of 
multilinear terms of odd degree in the $S_k$'s;
\item $f_M(S_{k-1}, S_{k-2},\ldots,S_{k-M})$ reduces to a single 
multilinear term containing an even number of $S_k$'s. 
\end{itemize}

For example, if $M=3$, the most general form for the memory function $f_3$,
satisfying the condition $<R_n>=0$, is either 
\begin{eqnarray*}
f_3(S_{k-1},S_{k-2},S_{k-3}) 
& = & a_1\eps_1S_{k-1}+a_2\eps_2S_{k-2} + a_3\eps_3S_{k-3}\\
&   &\qquad + a_{123}\eps_{123}S_{k-1}S_{k-2}S_{k-3},
\end{eqnarray*}
or
$$
f_3(S_{k-1},S_{k-2},S_{k-3}) = a_{ij}\eps_{ij}S_{k-i}S_{k-j},
$$
with $1\leq i\leq 3$ and $1\leq j\leq 3$.

In the sequel we will only consider two cases: either $f_M$ is a linear 
function of the $M$ previous steps, or the product of the $M$ previous steps.

Let us briefly examine these two cases. 

In what follows $(X_k)$ denotes an infinite sequence of \textit{iid} 
Bernoulli random variables such that, for all positive integers $k$, we have
$$ 
P(X_k=1) = P(X_k=-1) = \frac{1}{2}, 
$$ 

\textbf{1.} \textit{$f_M$ is linear.} If $(S_k)$ is an infinite sequence of 
random variables defined by 
\begin{equation} 
S_k = 
\begin{cases}
X_k & \text{if $k\le M$},\\[0.1cm]
X_k & \text{with probability $p$, if $k>M$},\\ 
\eps_j S_{k-j} & \text{with probability $qa_j$, if $k>M$,}
\end{cases}
\label{general-linear}
\end{equation}
where, for $j=1,2,\ldots,M$, the $\eps_j$ have fixed values chosen 
\textit{a priori\/} equal either to $1$ or $-1$, and the $a_j$ are $M$ 
nonnegative real numbers such that 
$$ 
a_1+a_2+\cdots+a_M = 1. 
$$ 

It is clear that, for $k\leq M$, $<S_k>=0$. Hence
\begin{eqnarray*}
<S_{M+1}> & = & p <X_{M+1}>+qa_1\eps_1<S_M>
+\cdots+qa_M\eps_M<S_1>\\
& = & 0\\
<S_{M+2}> & = & p <X_{M+2}>+qa_1\eps_1<S_{M+1}>
+\cdots+qa_M\eps_M<S_2>\\
& = & 0\\
&   & \cdots\\
<S_{M+\ell}> & = & p <X_{M+\ell}>+qa_1\eps_1<S_{M+\ell-1}>
+\cdots+qa_M\eps_M <S_\ell>\\
& = & 0,
\end{eqnarray*}
and, therefore, $<R_n>=0$, for all positive integers $n$.

\textbf{2.} \textit{$f_M$ is a single product of $r$ terms.} If $(S_k)$ is 
an infinite sequence of random variables defined by 
\begin{equation}
S_k = 
\begin{cases}
X_k & \text{if $k\le M$},\\[0.1cm]
X_k & \text{with probability $p$ if $k>M$},\\ 
\eps_{j_1j_2\cdots j_r}S_{k-j_1}S_{k-j_2}\cdots S_{k-jr} 
& \text{with probability $q$ if $k>M$}, 
\end{cases}
\label{general-product}
\end{equation} 
where $q=1-p$, $1\leq j_1<j_2<\cdots<j_r\leq M$, and 
$\eps_{j_1j_2\cdots j_r}$ is given and equal to either $1$ or $-1$.
Here again, for $k\leq M$, $<S_k>=0$. Hence, assuming first that $r=M$, 
we have
\begin{eqnarray*}
<S_{M+1}>& = & p<X_{M+1}>+q\eps_{12\cdots M}<S_1S_2\cdots S_M>\\
         & = & p<X_{M+1}>+q\eps_{12\cdots M}<S_1><S_2>\cdots <S_M>\\
         & = & 0\\
<S_{M+2}>& = & p<X_{M+2}>+q\eps_{12\cdots M}
                        <S_2S_3\cdots S_M S_{M+1}>\\
         & = & pq\eps_{12\cdots M}<S_2S_3\cdots S_M X_{M+1}>\\
         &   & \qquad + q^2\eps_{12\cdots M}<S_1S_2^2S_3^2\cdots S_M^2>\\
         & = & q^2\eps_{12\cdots M}<S_1>\;=\;0
\end{eqnarray*}
where in the expression of $S_{M+2}$ we have replaced $S_{M+1}$ by its
expression. More generally,
\begin{eqnarray*}
<S_{M+\ell}> & = & p<X_{M+\ell}>+q<S_\ell S_{\ell+1}\cdots S_{M+\ell-1}>\\
             & = & pq<S_\ell S_{\ell+1}\cdots X_{M+\ell-1}>+
            q^2<S_{\ell-1}S_\ell^2 S_{\ell+1}^2\cdots S_{M+\ell-2}^2>\\
             & = & q^2<S_{\ell-1}>,
\end{eqnarray*}
which shows that for all positive integers $k$, $<S_k>=0$, 
and, consequently, $<R_n>=0$.

Along similar lines, it can be shown that the result  $<R_n>=0$ remains 
valid for $r<M$, although its proof requires a larger number of substitutions. 

In the following sections we will show that, for any finite value of $M$, 
we can determine exactly the probability distribution of $<R_n>$, 
using a transfer matrix method. It will turn out that the rank of the matrix
is equal to $2^M$. Due to the increasing complexity of the eigensystem problem
with increasing $M$, this method becomes rapidly cumbersome, although it 
rather easily provides a numerical solution if the number of steps $n$ is
not too large. In order to illustrate the method, we will first briefly review
the Goldstein's solution of the Taylor's model. 

Since the asymptotic behavior of the variance $<R_n^2>$ is of particular 
interest to natural scientists, 
we will directly derive its exact expression through a recursive method 
that applies at any finite space dimensionality. From this expression we 
will be able to show that the asymptotic behavior $<R_n^2>$, 
when $n$ goes to infinity, exhibits a crossover phenomenon. 

\section{Taylor's model.}

The Taylor's model corresponds to Model~\ref{general-linear} for $M=1$, 
that is,
$$
S_1 = X_1,
$$
and, for all $k>1$,
$$
S_k=
\begin{cases}
X_k, & \text{with probability $p$}\\
\eps S_{k-1}, & \text{with probability $q=1-p$,}
\end{cases}
$$
where $(X_k)$ is a sequence of \textit{iid} Bernoulli variables such that, 
for all positive integers $k$, 
$$
P(X_k=-1) = P(X_k=1) = \frac{1}{2},
$$
and $\eps$ is a \textit{non-random number} equal to $\pm 1$.

If $p_n(k)$ denotes the probability that the walker will be at site $k$ after 
the $n$th step has been completed, we have
$$
p_n(k) = p_n^+(k)+p_n^-(k),
$$
where the superscript $\pm$ refers to the sign of the $n$-th step, that is
$$
p^{\pm}(k,n) = P((R_n=k)\cap (S_n = \pm 1)).
$$ 
Taking into account the conditional probabilities:
\begin{eqnarray*}
P(S_n=1\mid S_{n-1}=1)  =  P(S_n=-1\mid S_{n-1}=-1) & = &
\frac{1}{2}\left(p+(1+\eps)q\right)\\
P(S_n=1\mid S_{n-1}=-1) = P(S_n=-1\mid S_{n-1}=1) & = &
\frac{1}{2}\left(p+(1-\eps)q\right),
\end{eqnarray*}
we obtain the following recursion relations
\begin{eqnarray}
p_n^+(k) & = & \frac{1}{2}\left(p+(1+\eps)q\right)p_{n-1}^+(k-1)+
            \frac{1}{2}\left(p+(1-\eps)q\right)p_{n-1}^-(k-1)\nonumber\\
         & = &  \frac{1}{2}\left(1+\eps q\right)p_{n-1}^+(k-1)+
            \frac{1}{2}\left(1-\eps q\right)p_{n-1}^-(k-1)\label{pplus}\\
p_n^-(k) & = & \frac{1}{2}\left(p+(1-\eps)q\right)p_{n-1}^+(k+1)+
             \frac{1}{2}\left(p+(1+\eps)q\right)p_{n-1}^-(k+1)\nonumber\\
         & = &  \frac{1}{2}\left(1-\eps q\right)p_{n-1}^+(k+1)+
            \frac{1}{2}\left(1+\eps q\right)p_{n-1}^-(k+1)\label{pminus},
\end{eqnarray}
or, in a more condensed form which will be useful when we consider 
the case $M>1$,
\begin{equation}
p_n^\sigma(k)=  \frac{1}{2}\left(1-\eps q\right)p_{n-1}^+(k-\sigma)+
              \frac{1}{2}\left(1+\eps q\right)p_{n-1}^-(k-\sigma),
\label{psigma}
\end{equation}
with $\sigma=\pm1$.
The probability $\alpha$ in Goldstein's notations coincides here with 
$ \frac{1}{2}\left(1+\eps q\right)$.

The generating function of the probability distribution of the 
random walk is defined by
$$
f_n(x) = f_n^+(x) + f_n^-(x),
$$
with
\begin{eqnarray*}
f_n^+(x) & = & \sum_k p_n^+(k)\; x^{k}\\
f_n^-(x) & = & \sum_k p_n^-(k)\; x^{k},
\end{eqnarray*}
where the summation index $k$ runs from $-n$ to $n$ with step 2. 
Let
$$
\mathbf{f}_n(x) = 
\left(\begin{array}{c}
f_n^+(x)\\[0.2cm]
f_n^-(x)
\end{array}\right)
$$
and 
$$
{\mathbf M}(x) = 
\left(\begin{array}{cc}
  \frac{1}{2}\left(1+\eps q\right)\; x 
&  \frac{1}{2}\left(1-\eps q\right) \; x\\
\\
  \frac{1}{2}\left(1-\eps q\right) \; \frac{1}{x} 
&   \frac{1}{2}\left(1+\eps q\right)\; \frac{1}{x}\\
\end{array}\right).
$$
Then
\begin{equation}
\mathbf{f}_n(x) = \mathbf{M}(x) \; \mathbf{f}_{n-1}(x).
\label{1fiter}
\end{equation}
Iterating this recursion relation, we obtain
\begin{equation}
\mathbf{f}_n(x) = \mathbf{M}^{n-1}(x) \; \mathbf{f}_1(x),
\label{recur-1fn}
\end{equation}
where
$$
\mathbf{f}_1(x) = 
\left(\begin{array}{c}
f_{1}^+(x)\\[0.2cm]
f_{1}^-(x)
\end{array}\right)= 
\left(\begin{array}{c}
\frac{x}{2}\\[0.2cm]
\frac{1}{2x}
\end{array}\right).
$$

Let $\mathbf{R}(x)$ be the transformation matrix such that 
$$
\mathbf{D}(x) = \mathbf{R}^{-1}(x) \; \mathbf{M}(x) \; \mathbf{R}(x)
=\left(\begin{array}{cc}\lambda_1(x) & 0\\[0.2cm]
                         0 & \lambda_2(x)
\end{array}\right),
$$
where $\lambda_1(x)$ and $\lambda_2(x)$ are the eigenvalues of $\mathbf{M}(x)$.
Then
$$
\mathbf{f}_n(x) = 
\mathbf{R}(x)\; \mathbf{D}^{n-1}(x)\; \mathbf{R}^{-1}(x)\; \mathbf{f}_1(x) 
$$

The above relation involves the eigenvalues and eigenvectors of 
$\mathbf{M}(x)$ which can be readily obtained. 
However, the exact expression 
for the generating function of $p_n(k)$ is cumbersome to manipulate, 
and does not easily yield the variance. Therefore, we resort to an 
alternative and more direct method, which allows for a recursive 
evaluation of $<R_n^2>$:
\begin{eqnarray*}
<R_n^2> & = & <(S_1+S_2+\cdots+S_n)^2>\\
        & = & <(S_1+S_2+\cdots+S_{n-1})^2> \\
        &  & \qquad + 2<S_n\;(S_1+S_2+\cdots+S_{n-1})> + <S_n^2>\\
        & = & <R_{n-1}^2> + 2<S_n\;(S_1+S_2+\cdots+S_{n-1})> + <S_n^2>.\\
\end{eqnarray*}
As will appear below, the correlation coefficient $<S_n S_{n-k}>$ does not 
depend explicitly upon $n$, thus we use the notation
$$
c_k = <S_n S_{n-k}>,
$$
and obtain
$$
<R_n^2> = <R_{n-1}^2> + 2 (c_1+c_2 + \cdots + c_{n-1}) + 1,
$$
where we have taken into account that $<S_n^2>=1$. Evaluating the $c_k$'s, 
we get
\begin{eqnarray*}
c_1 & = & <S_nS_{n-1}> = p\;<X_nS_{n-1}> + \;\eps q\;<S_{n-1}^2>
=\eps q\\ 
c_2 & = & <S_nS_{n-2}> = p\;<X_nS_{n-2}> + \; \eps q\;<S_{n-1}S_{n-2}>
=\eps qc_1=q^2\\
    & \cdots & \\
c_k & = & <S_nS_{n-k}> = p\;<X_nS_{n-k}> + \eps q\;<S_{n-1}S_{n-k}>
=\eps qc_{k-1}=(\eps q)^k.
\end{eqnarray*}
Substituting the $c_k$'s in the expression of $<R_n^2>$ we obtain
\begin{eqnarray*}
<R_n^2> & = & <R_{n-1}^2> +\; 2\;(\eps q+q^2+\cdots
+(\eps q)^{n-1}) + 1\\
        & = & <R_{n-1}^2> +
\; 2\;\frac{\eps q-(\eps q)^n}{1-\eps q}+1,\\
\end{eqnarray*}
and iteration of this recursion relation yields
\begin{eqnarray}
<R_n^2> & = & n + \frac{2}{(1-\eps q)^2}
\;\left((n-1)\eps q - nq^2+(\eps q)^{n+1}\right)\nonumber\\
        & = & n + \frac{2}{(1-\eps (1-p))^2}\times\nonumber\\
        &   & \quad \left((n-1)\eps (1-p) - n(1-p)^2
+(\eps)^{n+1}(1-p)^{n+1}\right).\label{varofp}
\end{eqnarray}

It is quite remarkable that for a similar random walk on a $d$-dimensional 
simple cubic lattice, the above result is preserved. Of course, in this case, 
the sequence of one-dimensional $X_k$'s is replaced by a sequence of 
$d$-dimensional random vectors ${\mathbf X}_k$'s, and each elementary step 
can be performed along each of the $d$ axes, either in the positive or the 
negative direction.
For any fixed nonzero value of $p$, when $n$ goes to infinity, the term 
$(1-p)^{n+1}$ goes to zero, and $<R^2_n>$ behaves as 
$$
\frac{1+\eps q}{1-\eps q}\;n,
$$
and the random walk is eventually Gaussian at very large $n$. 
However, when $q$ goes to 1, according to whether $\eps$ is equal to 
$+1$ or $-1$, the prefactor of $n$ goes, respectively, either to infinity 
or to zero. Consequently, if $p$ is very small, the asymptotic behavior 
of $<R_n^2>$ should be carefully analyzed, distinguishing the two cases 
$\eps=1$ and $\eps=-1$.

\subsubsection{Persistent random walk ($\eps=1$).} 
\begin{figure}[h]
\centerline{\includegraphics*[scale=0.8]{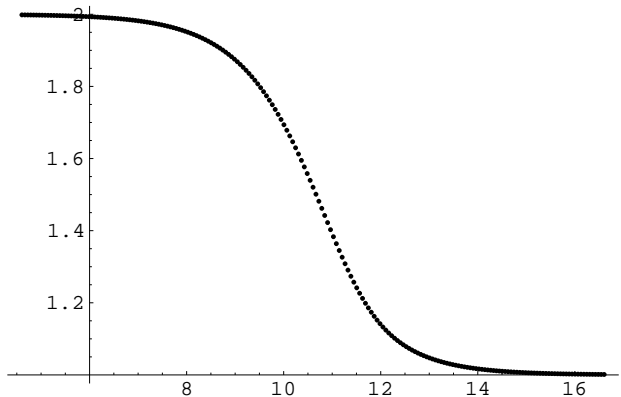}}
\caption{\label{fig:M1H5m5+}\textit{Exponent $E(n,p)$ of the persistent walk 
as a function of $\log n$, for $p=5\,10^{-5}$ and $n_{\max}=1.6\,10^7$.}}
\end{figure}

Replacing $\eps$ by $+1$ in 
\ref{varofp}, and expanding the resulting expression in powers of $p$, 
we obtain
$$
<R_n^2> = n + \frac{2}{p^2}\left(\frac{n^2p^2}{2}-\frac{np^2}{2}
-\frac{n(n^2-1)p^3}{6}+O(p^4)\right),
$$
which shows that, as expected, when $p$ tends to zero, $<R_n^2>$ tends to
$n^2$. Therefore, for a small fixed value of $p$, when $n$ increases, 
we expect, for the variance, a crossover from a $n^2$ behavior to an $n$ 
behavior. This crossover is controlled by the term $(1-p)^{n+1}$ in 
\ref{varofp}, and the above expansion of $<R_n^2>$ shows that the relevant 
parameter is the variable $a=n p$. To characterize this behavior, let 
define the exponent
$$
E(n,p) = \frac{\partial\log<R_n^2>}{\partial\log n},
$$
which is twice the Hurst exponent. A simple calculation yields
$$
\lim_{p\to 0} E\left(\frac{a}{p}, p\right) 
= \frac{a(1-\e^{-a})}{a+\e^{-a}-1}.
$$ 
This expression shows that, if $n$ is large but small compared to $1/p$ (\ie, 
$a$ is small), the exponent $E$ tends to 2, whereas if $n$ is large compared to 
$1/p$ ($a$ is large), $E$ tends to 1. This result is illustrated in
Figure~\ref{fig:M1H5m5+} for $p=0.00005$. Note that the crossover takes 
place around $n=\e^{11}\approx 6\,10^4$, $np$ being of the order of unity.

\begin{figure}[h]
\noindent
\begin{minipage}{0.46\linewidth}
\centering\epsfig{figure=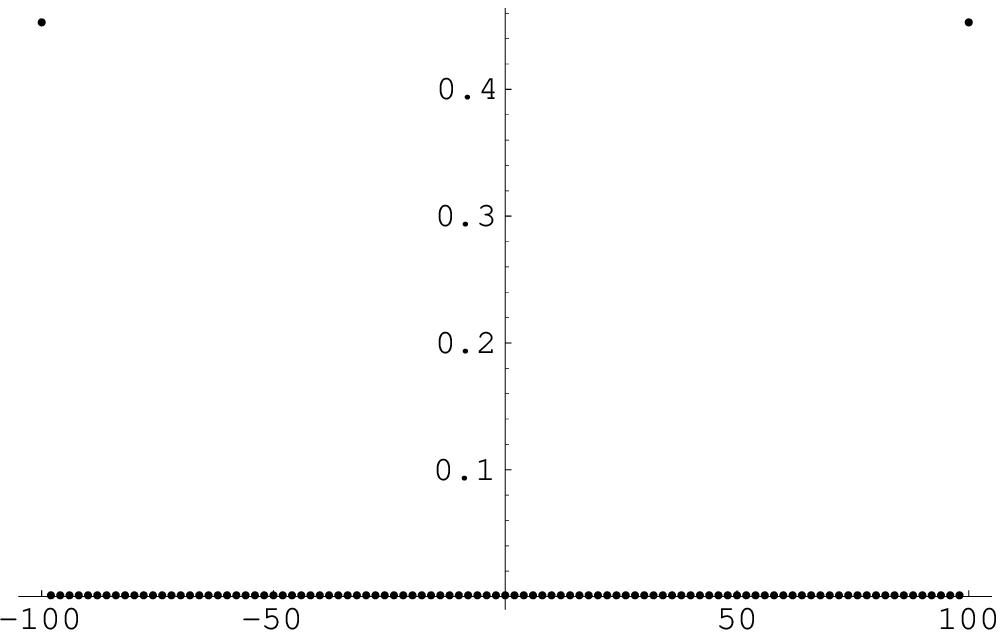, width=\linewidth}
\caption{\label{fig:M1n100p2m3+}\textit{$p_n(k)$ for $M=1$, 
$\eps=1$, $p=0.002$ and $n=100$.}}
\end{minipage}\hfill
\begin{minipage}{0.46\linewidth}
\centering\epsfig{figure=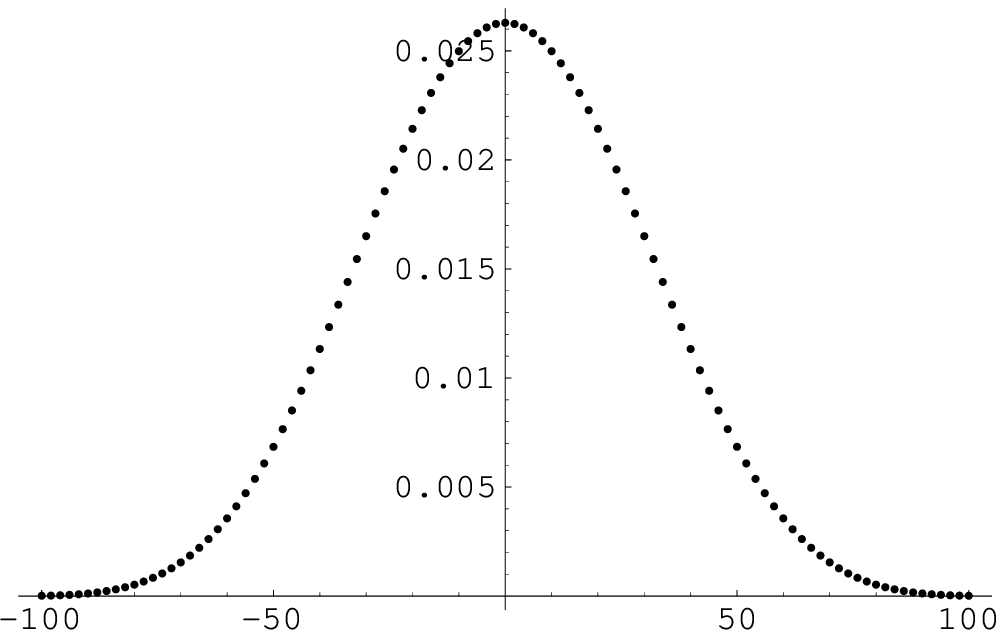, width=\linewidth}
\caption{\label{fig:M1n100p2m1+}\textit{$p_n(k)$ for $M=1$, 
$\eps=1$, $p=0.2$ and $n=100$.}}
\end{minipage}
\end{figure}

The crossover from $E(n,p)=2$ to $E(n,p)=1$, clearly observed in the above 
figure, where $p$ has a small fixed value and $n$ varies, can also be evidenced 
from the figures representing 
the probability distribution for a fixed value of $p$: 

\begin{itemize}
\item Figure~~\ref{fig:M1n100p2m3+}: at small $p$ ($np\ll 1$) most of the 
weight of the distribution is concentrated at the edges of the distribution range.
\item Figure~\ref{fig:M1n100p2m1+}: at large $p$ ($np\gg 1$) the significant 
weight of the distribution spreads out in a Gaussian-like profile, with a 
width roughly proportional to $\sqrt{n}$.
\end{itemize}

\vspace{0.3cm}

The fulfillment of both conditions (with perhaps the possibility that, in 
the case corresponding to $np\ll 1$, the localization takes place at various 
points of the distribution range, and not necessarily at the edges, 
as, for example, for the antipersistent walk discussed below) will be from 
now on considered as a sufficient condition for the occurrence of a crossover.

\subsubsection{Antipersistent random walk ($\eps=-1$).} Clearly, if 
$p=0$ the walk will be a zigzag 
walk, and $<R_n^2>=0$ (resp. $<R_n^2>=1$) if $n$ is even (resp. odd). If $p$ 
is small, a calculation similar to the previous one shows that:

$\bullet$ If $n p\gg 1$, $<R_n^2>$ behaves as $\frac{1}{2}n p$, and the 
oscillatory terms related to the parity of $n$ are negligible.

$\bullet$ If $n p\ll 1$, we find that $<R_n^2>$ behaves as $n p$ when $n$ is 
even, and as $1$ when $n$ is odd. Both expressions are the order of unity or 
less, so that $<R_n^2>$ behaves as $n^0$. 

\begin{figure}[h]
\centerline{\includegraphics*[scale=0.8]{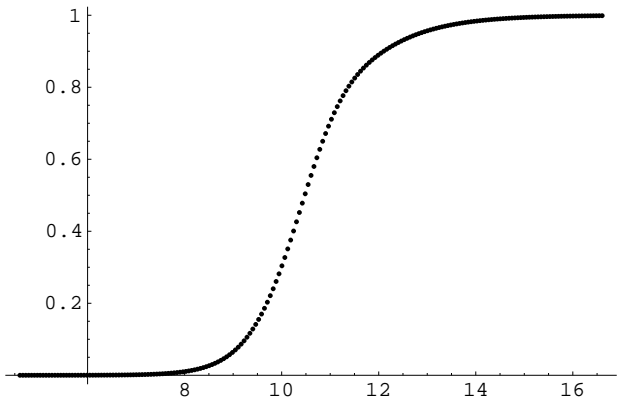}}
\caption{\label{fig:M1H5m5-}\textit{Exponent $E(n,p)$ of the antipersistent walk 
as a function of $\log n$ ($n$ odd),
for $p=5\,10^{-5}$ and $n_{\max}=1.6\,10^7$.}}
\end{figure}

Figure\ref{fig:M1H5m5-} represents the exponent characterizing the behavior 
of the variance. In order to avoid, for values of $n$ such that $np\ll 1$, 
the spurious behavior of $<R_n^2>$ in $n$ for even values of $n$, we have 
only considered the odd values of $n$.

\section{Linear memory function.}

We first address explicitly the case $M=2$, as a significant illustration
of the complexity which manifests as soon as $M>1$. Even in this case, a closed 
form for the probability distribution is  already difficult to write
down. However, we provide a formalism that can be fruitfully applied, at least 
for numerical purposes, to larger values of $M$. 

\subsection{Probability distribution for $M=2$.}

Let $p^{\sigma_1\sigma_2}(k,n)$ denote the probability that, after
$n$ steps, the random walker is at site $k$ and the steps $S_{n-1}$ 
and $S_n$ are , respectively, equal to $\sigma_1$ and $\sigma_2$, that is 
$$
p_n^{\sigma_1\sigma_2}(k)=P((R_n=k)\cap ((S_{n-1}=\sigma_1)\cap(S_n=\sigma_2))).
$$
If $S_{n-1}=\sigma_1$ and $S_n=\sigma_2$, then $R_{n-1}=k-\sigma_2$, and  
$p_n^{\sigma_1\sigma_2}(k)$ is a linear combination of two probabilities
of the form $p_{n-1}^{\sigma\sigma_1}(k-\sigma_2)$ where $\sigma=\pm 1$:
\begin{eqnarray}
p_n^{\sigma_1\sigma_2}(k) & = & \frac{1}{2}
\left(1+q(a_1\sigma_1\sigma_2\eps_1+a_2\sigma_2\eps_2)\right)
p_{n-1}^{+\sigma_1}(k-\sigma_2)\nonumber\\
&  & \quad + \frac{1}{2}
\left(1+q(a_1\sigma_1\sigma_2\eps_1-a_2\sigma_2\eps_2)\right)
p_{n-1}^{-\sigma_1}(k-\sigma_2).
\label{p12}
\end{eqnarray}
To this equation we associate the equation expressing 
$p_n^{\sigma_1{\overline{\sigma_2}}}(k)$, where ${\overline\sigma_2}=-\sigma_2$, 
since this probability is a linear combination of the two probabilities
$p_{n-1}^{\sigma\sigma_1}(k-\overline{\sigma_2}) =
p_{n-1}^{\sigma\sigma_1}(k+\sigma_2)$ where $\sigma=1$ or $-1$.
We have
\begin{eqnarray}
p_n^{\sigma_1{\overline\sigma_2}}(k) & = & \frac{1}{2}
\left(1+q(a_1\sigma_1{\overline\sigma_2}\eps_1
+a_2{\overline\sigma_2}\eps_2)\right)
p_{n-1}^{+\sigma_1}(k-{\overline\sigma_2})\nonumber\\
&  & \quad + \frac{1}{2}
\left(1+q(a_1\sigma_1{\overline\sigma_2}\eps_1
-a_2{\overline\sigma_2}\eps_2)\right)
p_{n-1}^{-\sigma_1}(k-{\overline\sigma_2}).
\label{p12bar}
\end{eqnarray}
The generating function of the probability distribution of $<R_n>$ is defined by
$$
f_n(x) = f_n^{++}(x) + f_n^{+-}(x) + f_n^{-+}(x) + f_n^{--}(x),
$$
with
$$
f_n^{\sigma_1\sigma_2}(x) =  \sum_k p_n^{\sigma_1\sigma_2}(k)\; x^{k},
$$
where the summation index $k$ runs from $-n$ to $n$ with step 2. 
From Equations~\ref{p12} and \ref{p12bar}, replacing $\sigma_1$ and 
$\sigma_2$ by all their possible respective values, we obtain
\begin{equation}
\left(\begin{array}{c}
f_n^{++}(x)\\[0.2cm]
f_n^{+-}(x)
\end{array}\right)
=
\mathbf{M^+}
\left(\begin{array}{c}
f_{n-1}^{++}(x)\\[0.2cm]
f_{n-1}^{-+}(x)
\end{array}\right),
\label{fMplus}
\end{equation}
where
\begin{equation}
\mathbf{M^+}(x) = \left(\begin{array}{cc}
\frac{1}{2}
\left(1+q(a_1\eps_1+a_2\eps_2)\right)\; x 
& \frac{1}{2}
\left(1+q(a_1\eps_1-a_2\eps_2)\right)\;x\\[0.2cm]
\frac{1}{2}
\left(1-q(a_1\eps_1
+a_2\eps_2)\right)\; \frac{1}{x} 
&\frac{1}{2}
\left(1-q(a_1\eps_1
-a_2\eps_2)\right)\;\frac{1}{x}
\end{array}\right),
\label{Mplus}
\end{equation}
and
\begin{equation}
\left(\begin{array}{c}
f_n^{-+}(x)\\[0.2cm]
f_n^{--}(x)
\end{array}\right)
=
\mathbf{M^-}
\left(\begin{array}{c}
f_{n-1}^{+-}(x)\\[0.2cm]
f_{n-1}^{-+}(x)
\end{array}\right),
\label{fMminus}
\end{equation}
where
\begin{equation}
\mathbf{M^-}(x) = \left(\begin{array}{cc}
\frac{1}{2}\left(1-q(a_1\eps_1-a_2\eps_2)\right)\; x 
&\frac{1}{2}\left(1-q(a_1\eps_1+a_2\eps_2)\right)\; x\\[0.2cm]
\frac{1}{2}\left(1+q(a_1\eps_1-a_2\eps_2)\right)\;\frac{1}{x} 
&\frac{1}{2}\left(1+q(a_1\eps_1+a_2\eps_2)\right)
\;\frac{1}{x}
\end{array}\right).
\label{Mminus}
\end{equation}

Equations~\ref{fMplus} and \ref{fMminus} can be grouped together in a unique
equation that reads
$$
\left(\begin{array}{c}
f_n^{++}(x)\\[0.18cm]
f_n^{+-}(x)\\[0.18cm]
f_n^{-+}(x)\\[0.18cm]
f_n^{--}(x)
\end{array}\right)
=
\left(\begin{array}{cc}
\mbox{\huge{$\mathbf{M^+}(x)$}} &  \mbox{\huge{$\mathbf{0}$}}\\[0.8cm]
\mbox{\huge{$\mathbf{0}$}} & \mbox{\huge{$\mathbf{M^-}(x)$}}
\end{array}\right)
\left(\begin{array}{c}
f_{n-1}^{++}(x)\\[0.18cm]
f_{n-1}^{-+}(x)\\[0.18cm]
f_{n-1}^{+-}(x)\\[0.18cm]
f_{n-1}^{--}(x)
\end{array}\right).
$$
Since the column vector on the right hand side of the above equation is not
correctly ordered, we have to introduce a $4\times 4$ permutation matrix, and
write this equation under the final form
\begin{equation}
\mathbf{f}_n(x)  =  \mathbf{M}(x)\;\mathcal{P}_4\; \mathbf{f}_{n-1}(x)
\label{2fiter}
\end{equation}
where 
$$
\mathbf{f}_n(x) = 
\left(\begin{array}{c}
f_n^{++}(x)\\[0.2cm]
f_n^{+-}(x)\\[0.2cm]
f_n^{-+}(x)\\[0.2cm]
f_n^{--}(x)\\
\end{array}\right),
$$

$$
\mathbf{M}(x) = 
\left(\begin{array}{cc}
\mathbf{M^+}(x) &  \mathbf{0}\\[0.2cm]
\mathbf{0} & \mathbf {M^-}(x)
\end{array}\right)
$$
and
$$
\mathcal{P}_4=
\left(\begin{array}{cccc}
1 & 0 & 0 & 0\\
0 & 0 & 1 & 0\\
0 & 1 & 0 & 0\\ 
0 & 0 & 0 & 1
\end{array}\right).
$$

Iterating Relation~\ref{2fiter}, we obtain
\begin{equation}
\mathbf{f}_n(x) 
= \left(\mathbf{M}(x)\;\mathcal{P}_4\right)^{n-2} \; \mathbf{f}_2(x),
\label{recur-2fn}
\end{equation}
where
$$
\mathbf{f}_2(x) = 
\left(\begin{array}{c}
f_{2}^{++}(x)\\[0.2cm]
f_{2}^{+-}(x)\\[0.2cm]
f_{2}^{-+}(x)\\[0.2cm]
f_{2}^{--}(x)\\
\end{array}\right)= 
\left(\begin{array}{c}
\frac{x^2}{4}\\[0.2cm]
\frac{1}{4}\\[0.2cm]
\frac{1}{4}\\[0.2cm]
\frac{1}{4x^2}
\end{array}\right).
$$
 
\vspace{0.3cm}
\begin{figure}[t]
\noindent
\begin{minipage}{0.46\linewidth}
\centering\epsfig{figure=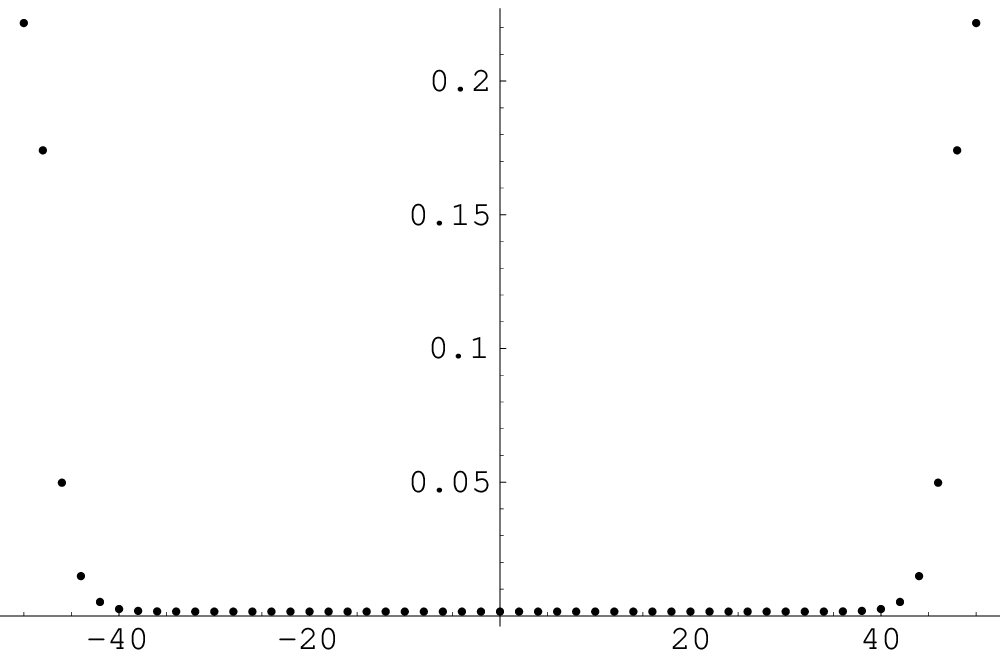, width=\linewidth}
\caption{\label{fig:LM2n50p5m3++}\textit{$p_n(k)$ for 
$M=2$, $a_1=0.5$, $a_2=0.5$, $\eps_1=1$, $\eps_2=1$, $p=0.005$, 
and $n=50$.}}
\end{minipage}\hfill
\begin{minipage}{0.46\linewidth}
\centering\epsfig{figure=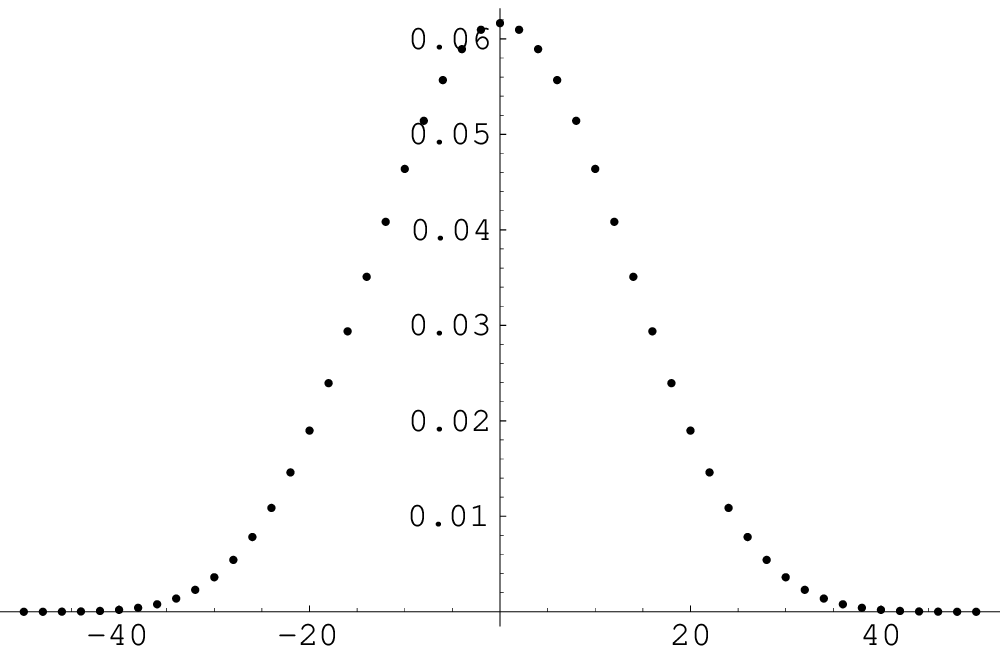, width=\linewidth}
\caption{\label{fig:LM2n50p5m1++}\textit{$p_n(k)$ for 
$M=2$, $a_1=0.5$, $a_2=0.5$, $\eps_1=1$, $\eps_2=1$, $p=0.5$, 
and $n=50$.}}
\end{minipage}
\end{figure}

Further progress toward the expression of the generating function $f_n(x)$
requires the explicit diagonalization of the $4\times 4$ matrix 
${\mathbf M}(x)\mathcal{P}_4$, followed by the procedure outlined in the $M=1$ 
case. The formal derivation of $f_n(x)$ is extremely complicated, and would 
reveal untractable for $M>2$ as will be seen below.
However, the formalism is well suited to numerically evaluate the probability
distribution of $R_n$ recursively using~\ref{recur-2fn}. 
Figures~\ref{fig:LM2n50p5m3++} and \ref{fig:LM2n50p5m1++} 
show the probability distribution $p_n(k)$ when $a_1=a_2=0.5$ and 
$\eps_1=\eps_2=1$ 
for two values of $p$ to illustrate cases $np\ll 1$ and $np\gg 1$.

Clearly, at the light of the comment at the end of subsection 3.0.1, the set 
of these two figures is a sure sign of the existence of a crossover.

\subsection{Variance for $M=2$.}

Correlations are more tedious to evaluate than in the case $M=1$. Let
$$
c_{1,\ell} = <S_\ell S_{\ell+1}>;
$$
then
\begin{eqnarray*}
c_{1,1} & = & <S_1 S_2> = 0\\
c_{1,2} & = & <S_2 S_3> = <S_2 (p X_3 + 
q(a_1\eps_1S_2 + a_2\eps_2 S_1))> = qa_1\eps_1
\end{eqnarray*}
and, for $\ell\ge 3$,
\begin{eqnarray*}
c_{1,\ell} & = & <S_\ell S_{\ell+1}> 
= <S_\ell (pX_{\ell+1} + qa_1\eps_1 S_\ell 
+ qa_2\eps_2 S_{\ell-1})> \\
        & = & qa_1\eps_1 + qa_2\eps_2 c_{1,\ell-1}.
\end{eqnarray*} 
Hence
\begin{equation}
c_{1,\ell} = qa_1\eps_1
\frac{1-(qa_2\eps_2)^{\ell-1}}{1-qa_2\eps_2}\quad(\ell\ge 1)
\label{correl-1ell}
\end{equation} 
Similarly, if 
$$
c_{2,\ell} = <S_\ell S_{\ell+2}>;
$$
then, for all positive integers $\ell$, 
\begin{eqnarray*}
c_{2,\ell} & = & <S_\ell S_{\ell+2}> 
= <S_\ell (pX_{\ell+2} + q(a_1\eps_1 S_{\ell+1} 
+ a_2\eps_2 S_\ell))>\\
        & = & qa_1\eps_1 c_{1,\ell} + qa_2\eps_2
\end{eqnarray*}
so that
\begin{eqnarray}
c_{2,\ell} = \frac{(qa_1\eps_1)^2}
{1-qa_2\eps_2}((1-qa_2\eps_2)^{\ell-1}) + qa_2\eps_2
\quad(\ell\ge 1).
\label{correl-2ell}
\end{eqnarray}
More generally, for $r>2$,
\begin{eqnarray}
c_{r,\ell} & = & <S_\ell S_{\ell+r}> = <S_\ell (p X_{\ell+r} + 
    q(a_1\eps_1 S_{k+r-1} + a_2\eps_2 S_{k+r-2}))>\nonumber\\
        & = & q(a_1\eps_1 c_{r-1,\ell} + a_2\eps_2 c_{r-2,\ell}).
\label{correl-rell}
\end{eqnarray}
From this recursion relation, we obtain the following expression of $c_{r,\ell}$ 
\begin{equation}
c_{r,\ell} = \frac{c_{1,\ell}\lambda_2-c_{2,\ell}}{\lambda_2-\lambda_1}
\lambda_1^{r-1}
+ \frac{c_{2,\ell}-c_{1,\ell}\lambda_1}{\lambda_2-\lambda_1}
\lambda_2^{r-1},\label{correlofr}
\end{equation}
where $\lambda_1$ and $\lambda_2$ are the roots of
$$
\lambda^2-qa_1\eps_1\lambda-qa_2\eps_2=0.
$$
It is easy to verify that $|\lambda_1|\le 1$ and $|\lambda_2|\le 1$, the equal 
sign being possible if $p=0$. For $p>0$, these conditions imply that the 
correlations $c_{r,\ell}$ decrease exponentially in $r$, and this, in turn, 
will ensure a linear dependence in $n$ for the variance.

Since the variance is given by
\begin{eqnarray*}
<R_n^2> & = & <(S_1+S_2+\cdots+S_n)^2>\nonumber\\
& = & n + 2\sum_{\ell=1}^{n-1}\sum_{r=1}^{n-\ell}c_{r,\ell},
\label{LMV}
\end{eqnarray*}
to find its its explicit expression, we need to evaluate the double summation. 
Summing first over $r$, we obtain
\begin{small}
\begin{eqnarray*}
\sum_{r=1}^{n-\ell}c_{r,\ell} & = &
\frac{1-\lambda_1^{n-\ell}}{(\lambda_2-\lambda_1)(1-\lambda_1)}
\left(\frac{qa_1\eps_1}{1-qa_2\eps_2}
(1-(qa_2\eps_2)^{\ell-1})(\lambda_2-qa_1\eps_1)-qa_2\eps_2\right)\\
& & + \frac{1-\lambda_2^{n-\ell}}{(\lambda_2-\lambda_1)(1-\lambda_2)}
\left(\frac{qa_1\eps_1}{1-qa_2\eps_2}
(1-(qa_2\eps_2)^{\ell-1})(qa_1\eps_1-\lambda_1)+qa_2\eps_2\right),
\end{eqnarray*}
\end{small}
where we have replaced $c_{2,\ell}$ by its expression given by \ref{correl-2ell}.
Summing now over $\ell$ finally yields
\begin{small} 
\begin{eqnarray*}
<R_n^2> & = & n + \frac{2(n-1)}{(\lambda_2-\lambda_1)(1-\lambda_1)}\left(
\frac{qa_1\eps_1(\lambda_2-qa_1\eps_1)}{1-qa_2\eps_2}-qa_2\eps_2
\right)\nonumber\\
& & - \frac{2}{(\lambda_2-\lambda_1)(1-\lambda_1)}\times
\frac{qa_1\eps_1(\lambda_2-qa_1\eps_1)}{1-qa_2\eps_2}\times
\frac{1-(qa_2\eps_2)^{n-1}}{1-qa_2\eps_2}\nonumber\\
& & -\frac{2}{(\lambda_2-\lambda_1)(1-\lambda_1)}\left(
\frac{qa_1\eps_1(\lambda_2-qa_1\eps_1)}{1-qa_2\eps_2}-qa_2\eps_2
\right)\frac{\lambda_1(1-\lambda_1^{n-1})}{1-\lambda_1}\nonumber\\
& & + \frac{2}{(\lambda_2-\lambda_1)(1-\lambda_1)}\times
\frac{qa_1\eps_1(\lambda_2-qa_1\eps_1)}{1-qa_2\eps_2}\times
\frac{\lambda_1(\lambda_1^{n-1}-(qa_2\eps_2)^{n-1})}{\lambda_1-qa_2\eps_2}
\nonumber\\
& & + \frac{2(n-1)}{(\lambda_2-\lambda_1)(1-\lambda_2)}\left(
\frac{qa_1\eps_1(qa_1\eps_1-\lambda_1)}{1-qa_2\eps_2}+qa_2\eps_2
\right)\nonumber\\
& & - \frac{2}{(\lambda_2-\lambda_1)(1-\lambda_2)}\times
\frac{qa_1\eps_1(qa_1\eps_1-\lambda_1)}{1-qa_2\eps_2}\times
\frac{1-(qa_2\eps_2)^{n-1}}{1-qa_2\eps_2}\nonumber\\
& & -\frac{2}{(\lambda_2-\lambda_1)(1-\lambda_2)}\left(
\frac{qa_1\eps_1(qa_1\eps_1-\lambda_1)}{1-qa_2\eps_2}+qa_2\eps_2
\right)\frac{\lambda_2(1-\lambda_2^{n-1})}{1-\lambda_2}\nonumber\\
& & + \frac{2}{(\lambda_2-\lambda_1)(1-\lambda_2)}\times
\frac{qa_1\eps_1(qa_1\eps_1-\lambda_1)}{1-qa_2\eps_2}\times
\frac{\lambda_2(\lambda_2^{n-1}-(qa_2\eps_2)^{n-1})}{\lambda_2-qa_2\eps_2}
\end{eqnarray*}
\end{small}

\begin{figure}[h]
\noindent
\begin{minipage}{0.46\linewidth}
\centering\epsfig{figure=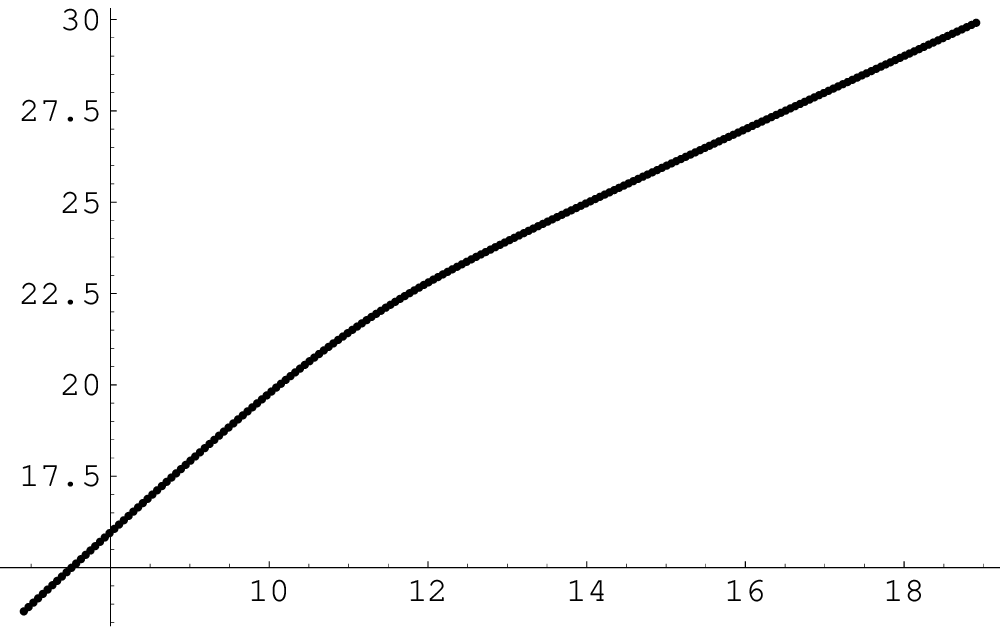, width=\linewidth}
\caption{\label{fig:PM2a5m1p5m5++}\textit{Log-log plot of $<R_n^2>$ vs.n 
for $M=2$, $\eps_1=1$, $\eps_2=1$, $a_1=a_2=0.5$, $p=5\,10^{-5}$, 
$n_{\max}=1.6\,10^8$.}}
\end{minipage}\hfill
\begin{minipage}{0.46\linewidth}
\centering\epsfig{figure=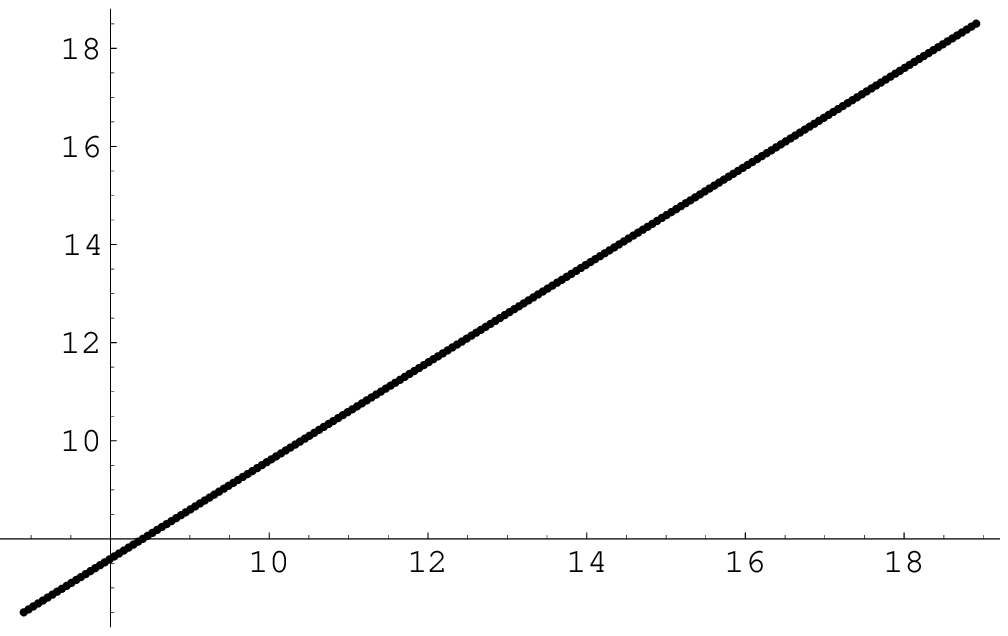, width=\linewidth}
\caption{\label{fig:PM2a5m1p5m5+-}\textit{Log-log plot of $<R_n^2>$ vs.n 
for $M=2$, $\eps_1=1$, $\eps_2=-1$, $a_1=a_2=0.5$, $p=5\,10^{-5}$, 
$n_{\max}=1.6\,10^8$.}}
\end{minipage}
\end{figure}

\begin{figure}[h]
\noindent
\begin{minipage}{0.46\linewidth}
\centering\epsfig{figure=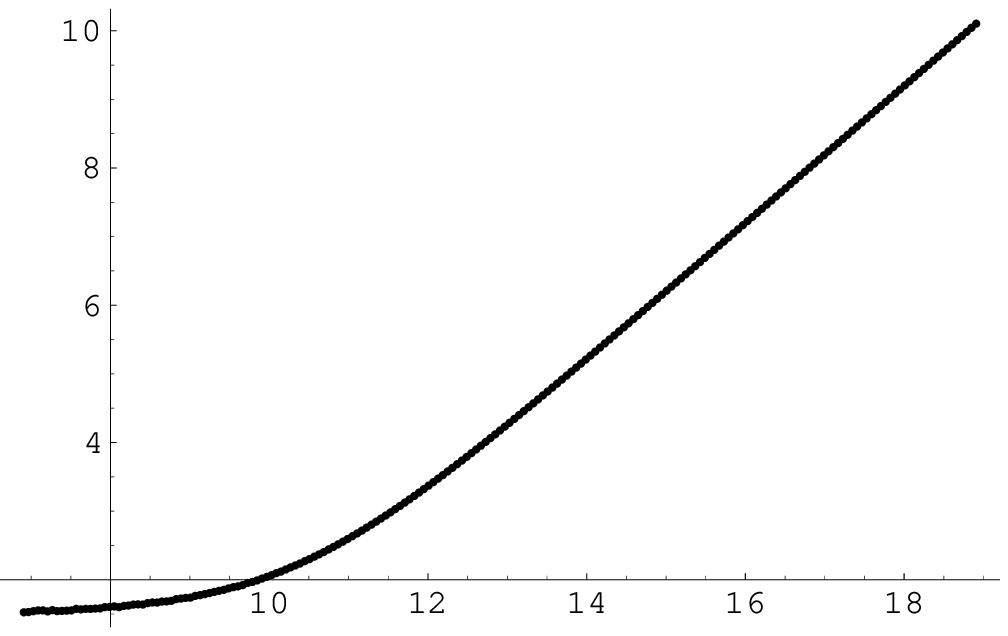, width=\linewidth}
\caption{\label{fig:PM2a5m1p5m5-+}\textit{Log-log plot of $<R_n^2>$ vs.n 
for $M=2$, $\eps_1=-1$, $\eps_2=1$, $a_1=a_2=0.5$, $p=5\,10^{-5}$, 
$n_{\max}=1.6\,10^8$.}}
\end{minipage}\hfill
\begin{minipage}{0.46\linewidth}
\centering\epsfig{figure=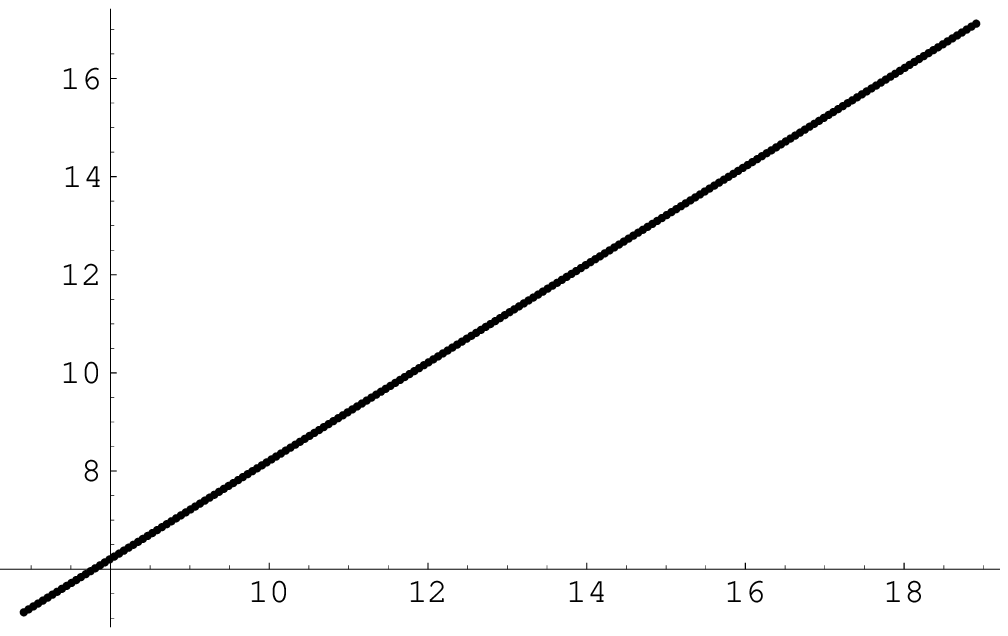, width=\linewidth}
\caption{\label{fig:PM2a5m1p5m5--}\textit{Log-log plot of $<R_n^2>$ vs.n 
for $M=2$, $\eps_1=-1$, $\eps_2=-1$, $a_1=a_2=0.5$, $p=5\,10^{-5}$, 
$n_{\max}=1.6\,10^8$.}}
\end{minipage}
\end{figure}

For our purpose, a complete analytic discussion of this expression 
would take us too far. Instead, we will illustrate the behavior of 
the variance with four typical choices for the set of the parameters 
$\eps_1,\eps_2,a_1,a_2$ in Figures~\ref{fig:PM2a5m1p5m5++},
\ref{fig:PM2a5m1p5m5+-}, \ref{fig:PM2a5m1p5m5-+}, and 
\ref{fig:PM2a5m1p5m5--}. Note that we obtain crossovers, respectively, 
similar to those obtained for the persistent and antipersistent walks 
of the Taylor's model when $\eps_2=1$ and $\eps_1$ is either equal 
to $1$ or $-1$. On the contrary when $\eps_2=-1$, There is no crossover, 
the variance behaves as $n$ for both signs of $\eps_1$.

\vspace{0.3cm}

\vspace{0.3cm}

\subsection{Linear memory for $M>2$.}

The matrix method used for the case $M=2$ is formally easy to generalize. It
yields, for the generating function, a recursion relation similar to 
\ref{2fiter}, where $\mathbf{f}_n$ is a $2^M$-component vector, 
$\mathbf{M}$ a $2^M\times 2^M$ matrix and $\mathcal{P}_4$ is replaced by a 
$2^M\times 2^M$ permutation matrix $\mathcal{P}_{2^M}$. With increasing values 
of $M$ it becomes rapidly untractable. Unfortunately, due to the large number 
of parameters when $M>2$, even the variance $<R_n^2>$ is unmanageable. 
This is not the case for the single product memory, as will be seen 
in the next section. However, there is, at least, one case for which we can 
predict that the variance exhibits the crossover observed in the persistent
walk of the Taylor's model. This occurs when all the $\eps_j$ 
($j=1,2,\ldots,M$) are equal to 1, since in this case, it is easy to verify
that for $p=0$, the variance behaves as $n^2$. 

\section{Single product memory function.}

Here again we will first consider the case $M=2$ as an illustration
of the complexity which manifests as soon as $M>1$. Since we will 
only consider the case of a single product, the notation 
$\eps_{j_1j_2\cdots j_r}$ will be replaced by $\eps$.

\subsection{Probability distribution for $M=2$.}

As in the linear memory function case, 
$p_n^{\sigma_1\sigma_2}(k)$ denotes the probability
$P((R_n=k)\cap((S_{n-1}=\sigma_1)\cap(S_n=\sigma_2)))$,
and as above, if $S_{n-1}=\sigma_1$ and $S_n=\sigma_2$, then 
$R_{n-1}=k-\sigma_2$, which implies that $p_n^{\sigma_1\sigma_2}(k)$ 
is a linear combination of the two probabilities  
$p_{n-1}^{\sigma\sigma_1}(k-\sigma_2)$, where $\sigma = \pm 1$.
Since, with a probability $q$, $S_n=\eps S_{n-1}S_{n-2}$, we have 
$\sigma_2=\eps\sigma\sigma_1$ or $\sigma=\eps\sigma_1\sigma_2$.
Therefore, using notations similar to those of the linear case, we have
$$
\left(\begin{array}{c}
f_n^{++}(x)\\[0.2cm]
f_n^{+-}(x)
\end{array}\right)
=
\mathbf{M^+}
\left(\begin{array}{c}
f_{n-1}^{++}(x)\\[0.2cm]
f_{n-1}^{-+}(x)
\end{array}\right),
$$
where
\begin{equation}
\mathbf{M^+}(x) = \left(\begin{array}{cc}
\frac{1}{2}
(1+q\eps)\; x 
& \frac{1}{2}
(1-q\eps)\;x\\[0.2cm]
\frac{1}{2}
(1-q\eps)\; \frac{1}{x} 
&\frac{1}{2}
(1+q\eps)\;\frac{1}{x}
\end{array}\right),
\label{PMplus}
\end{equation}
and
$$
\left(\begin{array}{c}
f_n^{-+}(x)\\[0.2cm]
f_n^{--}(x)
\end{array}\right)
=
\mathbf{M^-}
\left(\begin{array}{c}
f_{n-1}^{+-}(x)\\[0.2cm]
f_{n-1}^{-+}(x)
\end{array}\right),
$$
where
\begin{equation}
\mathbf{M^-}(x) = \left(\begin{array}{cc}
\frac{1}{2}
(1-q\eps)\; x 
& \frac{1}{2}
(1+q\eps)\;x\\[0.2cm]
\frac{1}{2}
(1+q\eps)\; \frac{1}{x} 
&\frac{1}{2}
(1-q\eps)\;\frac{1}{x}
\end{array}\right).
\label{PMminus}
\end{equation}

The above equations can be grouped together in a unique
equation that reads
$$
\left(\begin{array}{c}
f_n^{++}(x)\\[0.18cm]
f_n^{+-}(x)\\[0.18cm]
f_n^{-+}(x)\\[0.18cm]
f_n^{--}(x)
\end{array}\right)
=
\left(\begin{array}{cc}
\mbox{\huge{$\mathbf{M^+}(x)$}} &  \mbox{\huge{$\mathbf{0}$}}\\[0.8cm]
\mbox{\huge{$\mathbf{0}$}} & \mbox{\huge{$\mathbf{M^-}(x)$}}
\end{array}\right)
\left(\begin{array}{c}
f_{n-1}^{++}(x)\\[0.18cm]
f_{n-1}^{-+}(x)\\[0.18cm]
f_{n-1}^{+-}(x)\\[0.18cm]
f_{n-1}^{--}(x)
\end{array}\right).
$$
Since the column vector on the right hand side of the above equation is not
correctly ordered, we have to introduce a $4\times 4$ permutation matrix, and
write this equation under the final form
\begin{equation}
\mathbf{f}_n(x)  =  \mathbf{M}(x)\;\mathcal{P}_4\; \mathbf{f}_{n-1}(x)
\label{P2fiter}
\end{equation}
where 
$$
\mathbf{f}_n(x) = 
\left(\begin{array}{c}
f_n^{++}(x)\\[0.2cm]
f_n^{+-}(x)\\[0.2cm]
f_n^{-+}(x)\\[0.2cm]
f_n^{--}(x)\\
\end{array}\right),
$$

$$
\mathbf{M}(x) = 
\left(\begin{array}{cc}
\mathbf{M^+}(x) &  \mathbf{0}\\[0.2cm]
\mathbf{0} & \mathbf {M^-}(x)
\end{array}\right)
$$
and
$$
\mathcal{P}_4=
\left(\begin{array}{cccc}
1 & 0 & 0 & 0\\
0 & 0 & 1 & 0\\
0 & 1 & 0 & 0\\ 
0 & 0 & 0 & 1
\end{array}\right).
$$

Iterating Relation~\ref{P2fiter}, we obtain
\begin{equation}
\mathbf{f}_n(x) 
= \left(\mathbf{M}(x)\;\mathcal{P}_4\right)^{n-2} \; \mathbf{f}_2(x),
\label{recur-P2fn}
\end{equation}
where
$$
\mathbf{f}_2(x) = 
\left(\begin{array}{c}
f_{2}^{++}(x)\\[0.2cm]
f_{2}^{+-}(x)\\[0.2cm]
f_{2}^{-+}(x)\\[0.2cm]
f_{2}^{--}(x)\\
\end{array}\right)= 
\left(\begin{array}{c}
\frac{x^2}{4}\\[0.2cm]
\frac{1}{4}\\[0.2cm]
\frac{1}{4}\\[0.2cm]
\frac{1}{4x^2}
\end{array}\right).
$$

\vspace{0.3cm}
\begin{figure}[b]
\noindent
\begin{minipage}{0.46\linewidth}
\centering\epsfig{figure=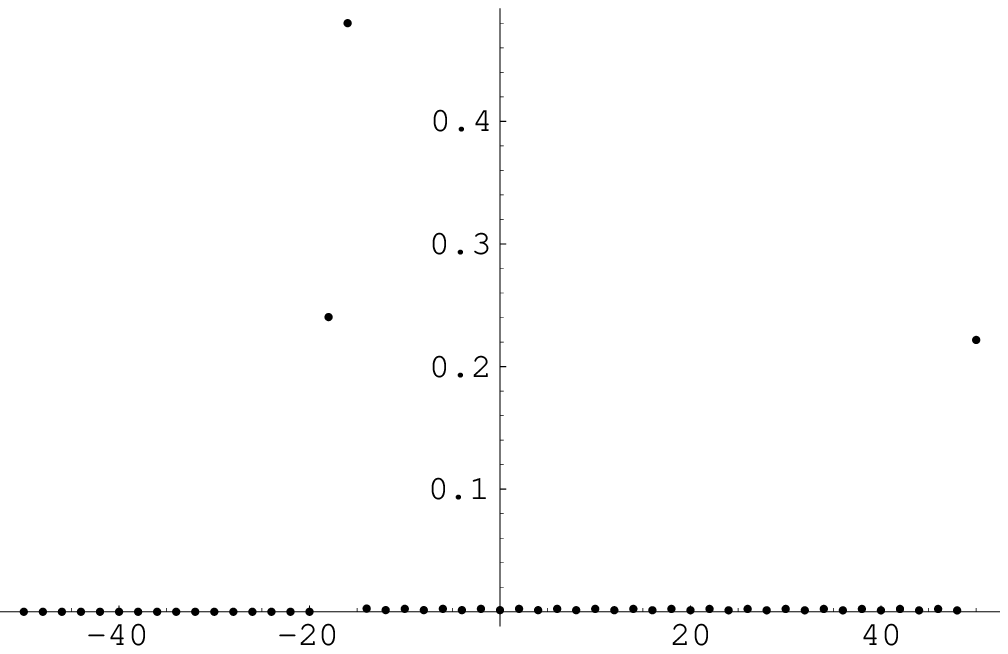, width=\linewidth}
\caption{\label{fig:PM2n50p5m3}\textit{$p_n(k)$ for 
$M=2$, $\eps=1$, $p=0.005$, and $n=50$.}}
\end{minipage}\hfill
\begin{minipage}{0.46\linewidth}
\centering\epsfig{figure=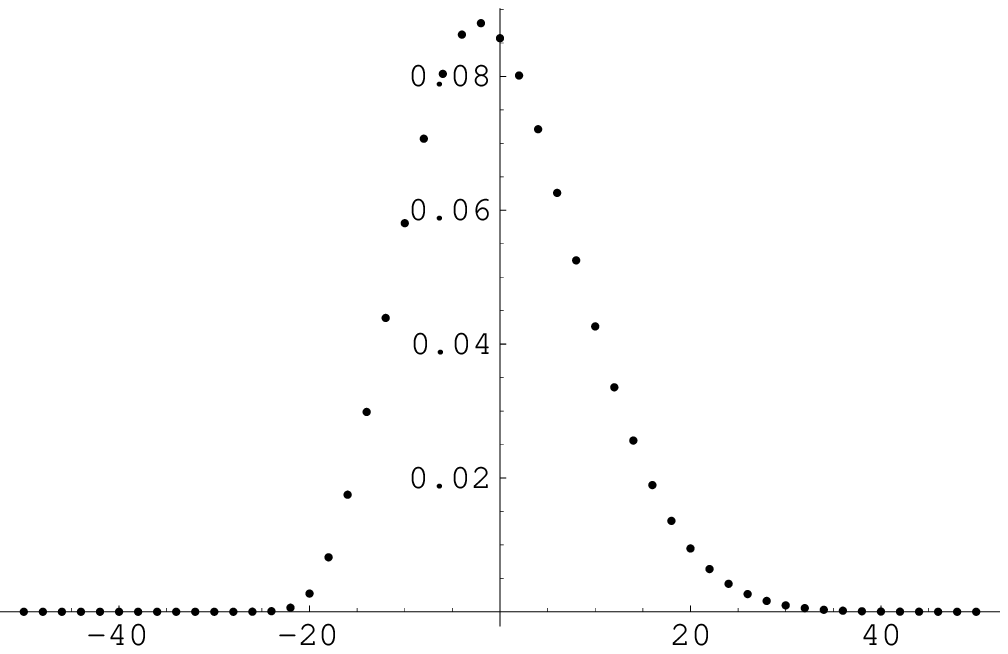, width=\linewidth}
\caption{\label{fig:PM2n50p5m1}\textit{$p_n(k)$ for 
$M=2$, $\eps=1$, $p=0.5$, and $n=50$.}}
\end{minipage}
\end{figure}

\vspace{0.3cm}
\begin{figure}[t]
\noindent
\begin{minipage}{0.46\linewidth}
\centering\epsfig{figure=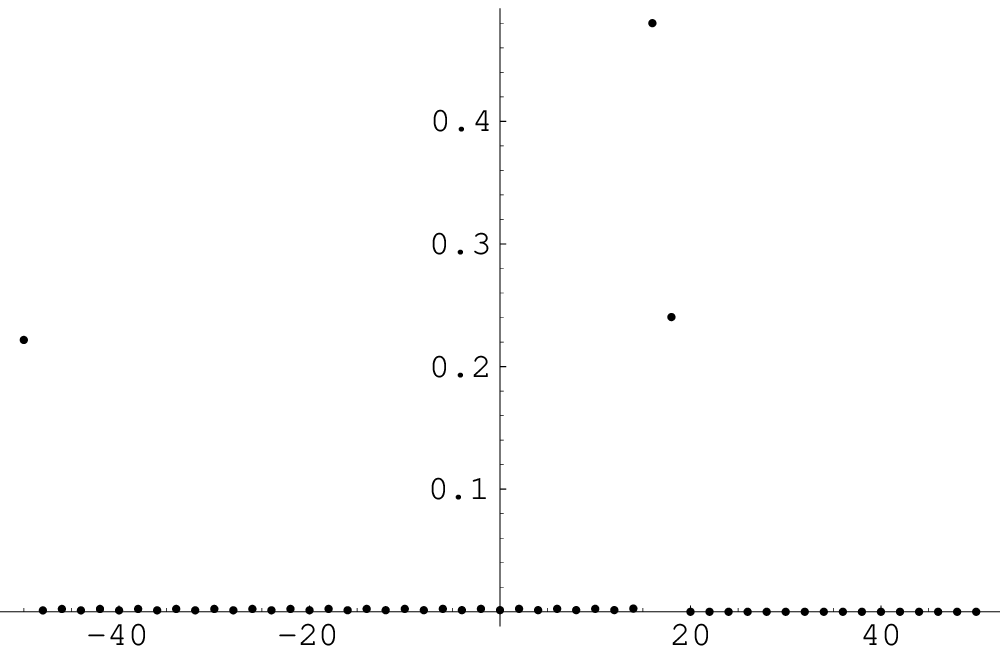, width=\linewidth}
\caption{\label{fig:PM2n50p5m3-}\textit{$p_n(k)$ for 
$M=2$, $\eps=-1$, $p=0.005$, and $n=50$.}}
\end{minipage}\hfill
\begin{minipage}{0.46\linewidth}
\centering\epsfig{figure=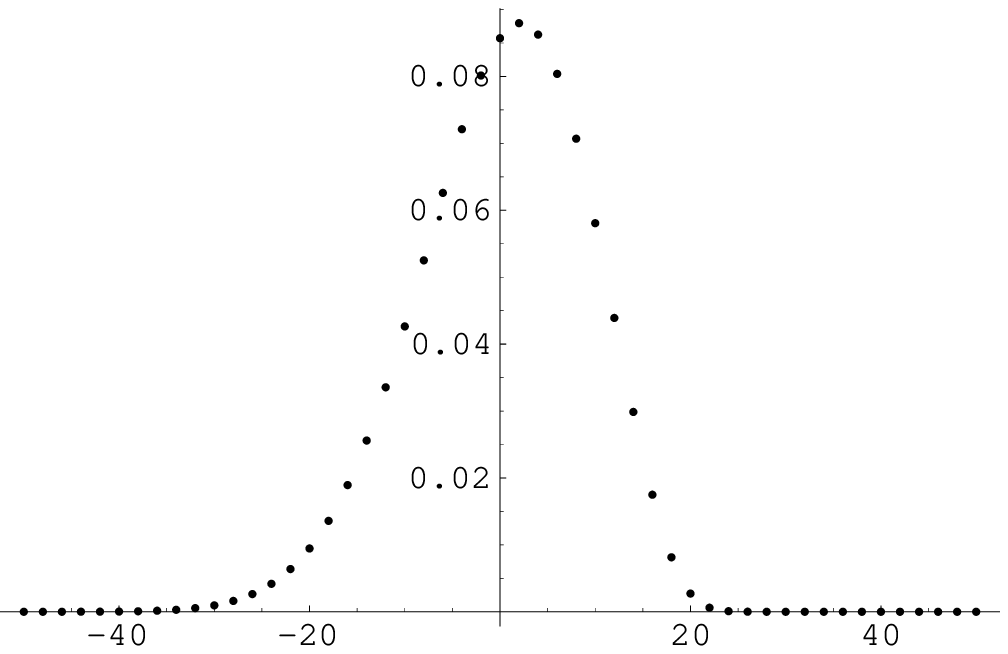, width=\linewidth}
\caption{\label{fig:PM2n50p5m1-}\textit{$p_n(k)$ for 
$M=2$, $\eps=-1$, $p=0.5$, and $n=50$.}}
\end{minipage}
\end{figure}

As in the linear case, further progress toward the expression of the generating 
function $f_n(x)$ is extremely complicated, and would 
reveal untractable for $M>2$.
However the formalism is well suited to numerically evaluate 
the probability distribution of $R_n$ recursively using~\ref{recur-P2fn}. 
Figures~\ref{fig:PM2n50p5m3}, \ref{fig:PM2n50p5m1},  \ref{fig:PM2n50p5m3-},
\ref{fig:PM2n50p5m1-} show the probability distribution $p_n(k)$ 
for two values of $p$ to illustrate cases $np\ll 1$ and $np\gg 1$ when
$\eps=\pm1$.

At the light of the comment at the end of subsection 3.0.1, 
Figures~\ref{fig:PM2n50p5m3}, \ref{fig:PM2n50p5m1}, \ref{fig:PM2n50p5m3-},
\ref{fig:PM2n50p5m1-} reveal the existence of crossovers. This property 
will be confirmed directly in the next subsection.

\subsection{Variance for $M=2$.}

In order to evaluate recursively the variance, we need to determine 
the correlations. We have
\begin{eqnarray*}
<S_\ell S_{\ell+1}> & = & \eps q<S_\ell (S_{\ell}S_{\ell-1})>\\
             & = & \eps q <S_{\ell-1}>=0;
\end{eqnarray*}
\begin{eqnarray*}
<S_\ell S_{\ell+2}> & = & \eps q<S_\ell (S_{\ell+1}S_\ell)>\\
             & = & \eps q <S_{\ell+1}>=0;
\end{eqnarray*}
\begin{eqnarray*}
<S_\ell S_{\ell+3}> & = & \eps q<S_\ell (S_{\ell+2}S_{\ell+1})>\\
             & = & \eps^2 q^2 <S_\ell((S_{\ell+1}S_\ell)S_{\ell+1}>\\
             & = & q^2;
\end{eqnarray*}
and, for $m>\ell+3$,
\begin{eqnarray*}
<S_\ell S_m> & = & \eps q<S_\ell (S_{m-1}S_{m-2})>\\
             & = & \eps^2 q^2 <S_\ell((S_{m-2}S_{m-3})S_{m-2}>\\
             & = & q^2 <S_\ell S_{m-3}>.
\end{eqnarray*}

\begin{figure}[h]
\centerline{\includegraphics*[scale=0.8]{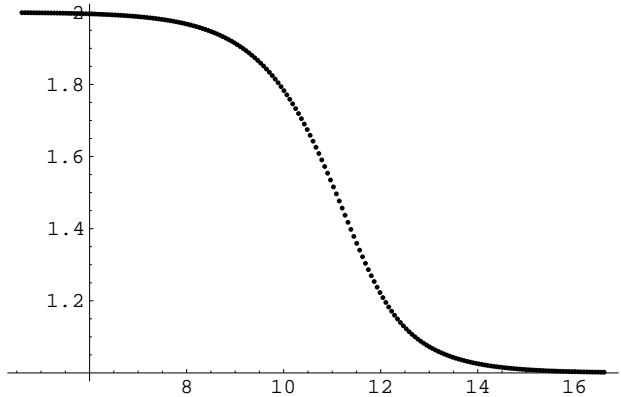}}
\caption{\label{fig:PM2H5m5}\textit{Exponent $E(n,p)$ as a function of 
$\log n$, for $M=2$, $p=5\,10^{-5}$ and $n_{\max}=1.6\,10^7$.}}
\end{figure}

These results provide all the correlations:
\begin{equation} 
<S_\ell S_m> = 
\begin{cases}
0 & \text{if $m-\ell$ is not a multiple of 3},\\[0.1cm]
q^{2(m-\ell)/3} & \text{if $m-\ell$ is a multiple of 3}. 
\end{cases}
\label{correlPM2}
\end{equation}
Note that the correlations do not depend upon $\eps$.
On the other hand, the variance satisfies the recursion relation
$$
<R_n^2> = <R_{n-1}^2> + 1 + 2<S_1S_n + S_2S_n + \cdots + S_{n-1}S_n>. 
$$
Iterating this relation and taking into account \ref{correlPM2}, we 
finally obtain
\begin{eqnarray}
<R_n^2> & = & n + (n-3)q^2+(n-6)q^4+\cdots+
(n-3\lfloor n/3\rfloor) q^{2\lfloor n/3\rfloor}\nonumber\\
& = & n + n q^2\frac{1-q^{2\lfloor n/3\rfloor}}{1-q^2}\nonumber\\
&  &\quad - 3q^2\frac{\lfloor n/3\rfloor q^{2\lfloor n/3\rfloor+2}-
\left(\lfloor n/3\rfloor+1\right)q^{2\lfloor n/3\rfloor} +1}
{(1-q^2)^2},\label{varPM2}
\end{eqnarray}
where $\lfloor x\rfloor$ denotes the largest integer less or equal to $x$.
On this expression of $<R_n^2>$ we readily verify that for $q=0$ the 
variance is equal to $n$, while for $q=1$ it takes the value
$$
n + (2n-3)\lfloor n/3\rfloor - 3 (\lfloor n/3\rfloor)^2 = \lceil n^2/3\rceil,
$$
where $\lceil x\rceil$ denotes the smallest integer greater or equal to $x$.
Since the functions floor and ceiling are not differentiable, we define the
crossover exponent $E(n,p)$ by
$$
E(n,p)= \frac{n}{2}\frac{<R_{n+1}^2>(p)-<R_{n-1}^2>(p)}{<R_n^2>(p)}.
$$
Figure~\ref{fig:PM2H5m5} represents the variation of $E(n,p)$ as a function of
of $\log n$ for $p=5\,10^{-5}$.

\vspace{0.3cm}

\subsection{Probability distribution for $M>2$.}

The matrix method used for the case $M=2$ is formally easy to generalize. It
yields, for the generating function, a recursion relation similar to 
\ref{P2fiter}, where $\mathbf{f}_n$ is a $2^M$-component vector, 
$\mathbf{M}$ a $2^M\times 2^M$ matrix and $\mathcal{P}_4$ is replaced by a 
$2^M\times 2^M$ permutation matrix $\mathcal{P}_{2^M}$. With increasing values 
of $M$ it becomes rapidly untractable. However, the variance $<R_n^2>$ can be
exactly evaluated as we will show in next subsection.

\subsection{Variance for $M>2$.}
 
In order to evaluate recursively the variance, we need to determine 
the correlations. By definition, $<S_\ell S_m>=0$ if $\ell<m\le M$. This is
also true if $m=M+1$:
\begin{eqnarray*}
<S_\ell S_{M+1}> & = & q\eps<S_\ell(S_1S_2\cdots S_\ell\cdots S_M)>\\
                 & = &q\eps <S_1><S_2>\cdots<S_M>=0,
\end{eqnarray*}
due the fact that the random variables $S_1, S_2, \ldots, S_M$ 
are independent, and $<S_\ell^2>=1$. 

Considering now the case: $<S_\ell S_m>$ (any $\ell$ and $m>\ell$), we first 
note that
\begin{eqnarray*}
<S_\ell S_{\ell+M+1}> & = & q\eps<S_\ell(S_{\ell+1}S_{\ell+2}\cdots S_{\ell+M})>\\
                 & = & (q\eps)^2<(S_\ell S_{\ell+1}\cdots S_{\ell+M-1})^2>\\
                 & = & q^2,
\end{eqnarray*}      
whereas, in the general case,
\begin{eqnarray}
<S_\ell S_{\ell+r}> & = & q\eps
<S_\ell(S_{\ell+r-M}S_{\ell+r-M+1}\cdots S_{\ell+r-1})>\nonumber\\
                 & = & q^2<S_\ell S_{\ell+r-(M+1)}>,
\label{correliterPM}
\end{eqnarray}   
where it is assumed that $\ell+r-(M+1)$ is positive but may be smaller 
than $\ell$. Iterated application of~\ref{correliterPM} in the presence 
of the initial conditions shows that, for 
$m-\ell = \big\lfloor{\frac{m-\ell}{M+1}}\big\rfloor + b$ ($0\leq b<M+1$),
\begin{equation}
<S_\ell S_m> = 
\begin{cases}
0 & \text{if $b>0$}\\
q^{2\big\lfloor\frac{m-\ell}{M+1}\big\rfloor} & \text{if $b=0$}\\
\end{cases}
\label{correlPMgeneral}
\end{equation}
Substitution of~\ref{correlPMgeneral} in the recursion relation satisfied by 
the variance:
$$
<R_n^2> = <R_{n-1}^2> + 1 + 2<S_1S_n + S_2S_n + \cdots + S_{n-1}S_n>. 
$$
yields
\begin{eqnarray}
<R_n^2> & = & n + (n-(M+1))q^2 + (n -2(M+1))q^4+\cdots\nonumber\\
& & +\left(n-\Big\lfloor\frac{n}{M+1}\Big\rfloor(M+1)\right)
q^{2\lfloor\frac{n}{M+1}\rfloor}\nonumber\\
& = & n + n q^2\frac{1-q^{2\lfloor\frac{n}{M+1}\rfloor}}{1-q^2}
-(M+1)q^2\nonumber\\
& & \times\frac{1 - \left(\big\lfloor\frac{n}{M+1}\big\rfloor+1\right)
q^{2\lfloor\frac{n}{M+1}\rfloor}+\big\lfloor\frac{n}{M+1}\big\rfloor
q^{2\lfloor\frac{n}{M+1}\rfloor+2}}{(1-q^2)^2},\label{varPMofq}
\end{eqnarray}
which is the generalization for any $M$ of the result~\ref{varPM2} 
obtained for $M=2$. Note that $M=1$ is not a particular case of~\ref{varPMofq}.

\begin{figure}[h]
\centerline{\includegraphics*[scale=0.8]{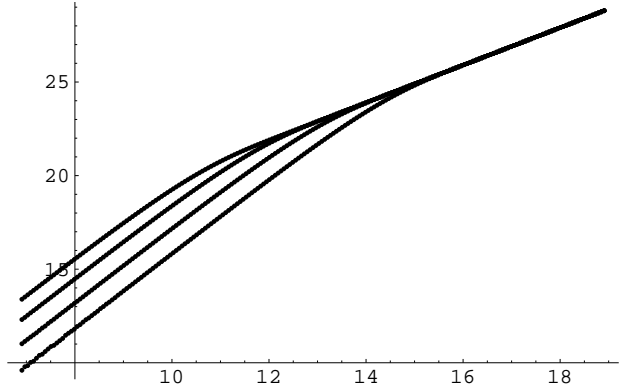}}
\caption{\label{fig:multiPMVp5m5}\textit{Log-log plot of the variance 
as a function of $n$, 
for $M=2,8,32,128$ (to be read downwards), $p=5\,10^{-5}$ and 
$n_{\max}=1.6\,10^8$.}}
\end{figure}

When $q=0$, $<R_n^2>=n$, whereas, when $p$ goes to zero ($q\to 1$), $<R_n^2>$ 
tends to
$$
n + (2n-(M+1))\bigg\lfloor\frac{n}{M+1}\bigg\rfloor,
$$
which is of the order of $\frac{n^2}{M+1}$ for large $n$.
These two limiting behaviors are the signal of a crossover, which is evidenced
in Figure~\ref{fig:multiPMVp5m5}. This figure shows that the variance is a 
decreasing function of $M$ at a given $n$.

\vspace{0.3cm}

\begin{figure}[h]
\noindent
\begin{minipage}{0.46\linewidth}
\centering\epsfig{figure=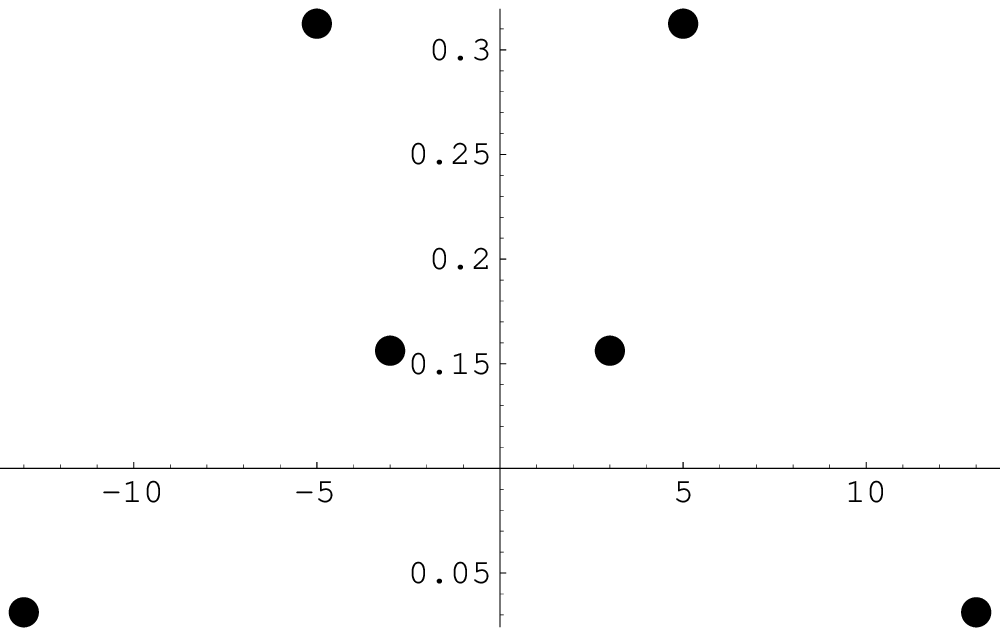, width=\linewidth}
\caption{\label{fig:PM5n13p0+}\textit{$p_n(k)$ for 
$M=5$, $\eps=1$, $p=0$, and $n=13$. $<R_{13}^2>=29$.}}
\end{minipage}\hfill
\begin{minipage}{0.46\linewidth}
\centering\epsfig{figure=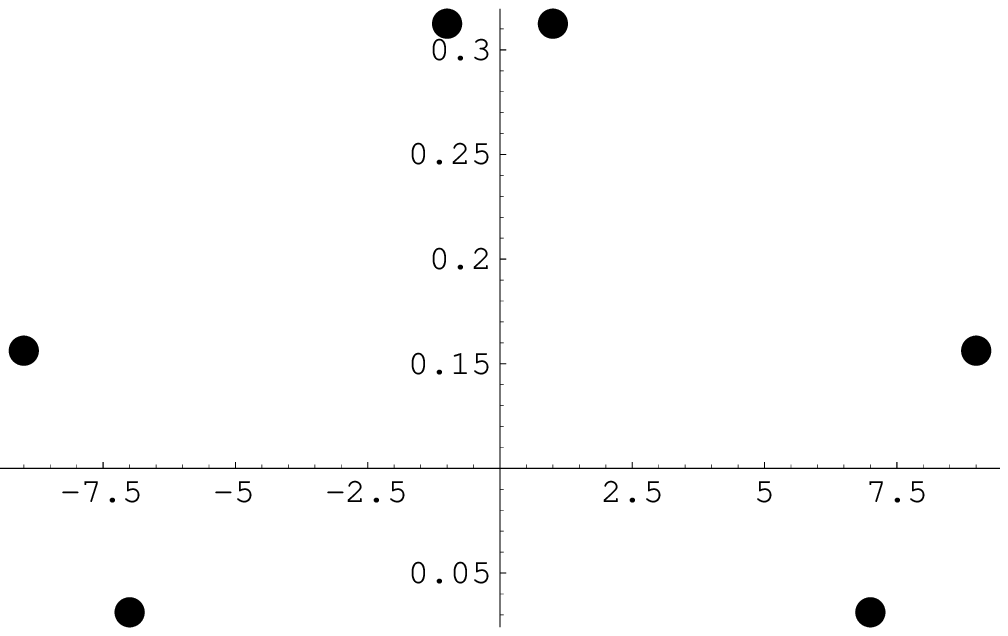, width=\linewidth}
\caption{\label{fig:PM5n13p0-}\textit{$p_n(k)$ for 
$M=5$, $\eps=-1$, $p=0$, and $n=13$. $<R_{13}^2>=29$.}}
\end{minipage}
\end{figure}

The general expression of the variance is an even function of $q$, and, 
therefore, does not depend upon $\eps$. Such is not the case for the 
probability distribution $p_n(k)$ of $R_n$ as in Figures~\ref{fig:PM5n13p0+} and
\ref{fig:PM5n13p0-}.

\vspace{0.3cm}

Here the probability distribution are symmetric. This is not always the case as
can be observed in Figures~\ref{fig:PM2n50p5m3}, \ref{fig:PM2n50p5m1},
\ref{fig:PM2n50p5m3-}, and \ref{fig:PM2n50p5m1-}.

\section{Conclusion and perspectives}

We have studied a family of correlated random walks with a finite memory
range. These walks are a generalization of the Taylor's walk.
We have established a matrix formalism for the generating functions of 
the probability distributions of these walks. We have also derived in many 
cases an analytic expression for the variance. As a function of the number 
of steps $n$, for each of these walks, the variance becomes ultimately
linear in $n$ in the limit $n\to\infty$. However, at comparatively small $n$,
a different behavior of the variance can be observed leading to the existence
of a crossover phenomenon. This feature shows that, when analyzing experimental 
results, one finds a Hurst exponent different from $\frac{1}{2}$, this does not 
necessarily imply an infinite-range memory. If, experimentally feasible, one 
should increase significantly the number of steps to cross-check the stability 
of the exponent before concluding that the system under consideration exhibits 
a non-Gaussian behavior.

The ultimate Gaussian behavior of the variance
at very large $n$ is a direct consequence of the exponential decay of the 
step-step correlations as a function of the absolute value of the difference 
between step indices. Therefore, in order to find a different behavior, one 
has to consider step-step correlations decreasing more slowly than 
exponentially. This is what may happen in the case of a power law decay
as can be shown as follows.
For a symmetric random walk, the variance is given by
\begin{eqnarray*}
<R_n^2> & = & n + 2 \sum_{\ell=1}^{n-1}\sum_{m=\ell+1}^n <S_\ell S_m>\\
& = & n + 2 \sum_{\ell=1}^{n-1}\sum_{r=1}^n <S_\ell S_{\ell+r}>\\
& = & n + 2 \sum_{\ell=1}^{n-1}\sum_{r=1}^n c_{r,\ell},
\end{eqnarray*}
where as above $c_{r,\ell}=<S_\ell S_{r+\ell}>$. If we now assume that, 
for some reason,
$$
c_{r,\ell} = \frac{1}{r^\alpha}
$$
independently of $\ell$, then
$$
<R_n^2> = n + 2\sum_{r=1}^{n-1}\frac{n-r}{r^\alpha}.
$$

The possible non-Gaussian behavior of $<R_n^2>$ can only result from the 
asymptotic behavior at large $n$ of the summation in the right hand side 
of the above relation. According to the value of $\alpha$, the 
asymptotic behaviors of
$$
S(n,\alpha) = \sum_{r=1}^{n-1}\frac{n-r}{r^\alpha}
$$
depends on $\alpha$ as follows:
\begin{itemize}
\item{if $\alpha>1$, $S(n,\alpha)\sim n\zeta(\alpha)$, where $\zeta$ is the Riemann 
function.}
\item{if $\alpha=1$, $S(n,1)\sim n\log n$.}
\item{if $\alpha<1$, $S(n,\alpha)\sim\frac{1}{(1-\alpha)(2-\alpha)}n^{2-\alpha}$.}
\end{itemize}
Note that the power law decay for the step-step correlations is only a 
sufficient condition to obtain a non-trivial Hurst exponent equal to $1-a/2$.

\section*{References}

{\parindent = 0pt 

[1] For a review of early work on random walk, see W. Feller, 
An Introduction to Probability Theory and Its Applications, 
John Wiley and Sons, New York, 1960, Volume I, Chapter XIV.

[2] J. J. Collins and C. J. De Luca, Random Walking during Quiet 
Standing, Phys. Rev. Lett., {\bf 73}, 764--767, 1994.

[3] L. Nunes Amaral, A. Goldberger, P. Ivanov and E. Stanley, Scale-Indepen\-dent
Measures and Pathologic Cardiac Dynamics, Phys. Rev. Lett., {\bf 81}, 
2388--2391, 1998.

[4] W. Li, The Study of Correlated Structures of DNA Sequences: A Critical
Review, lanl archives, adap-org/9704003

[5] C.-K. Peng, S. Buldyrev, A. Goldberger, S. Havlin, F. Sciortino, 
M. Simons and E. Stanley, Long-range correlations in nucleotide sequences,
Nature, {\bf 356}, 168--170, 1992.

[6] G. Abramson, P. Alemany and H. Cerdeira, Noisy L\'evy walk analog
of two-dimensional DNA walks for chromosomes of S. cerevisiae, Phys. Rev. E 
{\bf 58,} 914--918, 1998.  

[7] L. Liebovitch and W. Yang, Transition from persistent to antipersistent 
correlation in biological systems, Phys. Rev. E, {\bf 56}, 4557--4566, 1997.

[8] P. Ivanov, L. Nunes Amaral, A. Goldberger and E. Stanley, Stochastic
feedback and the regulation of biological rhythms, Europhys. Lett., {\bf 43}, 
363--368, 1998. 

[9] R. Mantegna and E. Stanley, Scaling behaviour in the dynamics of an 
economic index, Nature, {\bf 376}, 46--49, 1995.

[10] R. Mantegna and E. Stanley, Turbulence and financial markets, Nature, 
{\bf 383}, 587--588, 1996.

[11] E. Scalas, Scaling in the market of futures, Physica A, {\bf 253}, 
394--402, 1998.

[12] G. I. Taylor, Diffusion by continuous movements, Proceedings 
of the London Mathematical Society, {\bf 20}, 196--212, 1921/22.

[13] S. Goldstein, On diffusion by discontinuous movements, and the
telegraph equation, The Quarterly Journal of Mechanics and Applied 
Mathematics {\bf 4}, 129--156, 1951.

[14] J. Gillis, Correlated random walk, Proc. Camb. Phys. Soc. 
{\bf 51}, 639--651, 1955.

[15] E. Renshaw and R. Henderson, The correlated random walk, Journal 
of Applied Probability, {\bf 18}, 403--414, 1981.

[16] A. Chen and E. Renshaw, The Gillis-Domb-Fisher correlated random
walk, Journal of Applied Probability, {\bf 29}, 792--813, 1992.

[17] G. H. Weiss, Aspects and Applications of the random walk, 
North-Holland, Amsterdam, 1994.

}

\end{document}